\documentclass{article}

\usepackage{arxiv}

\usepackage[utf8]{inputenc} % allow utf-8 input
\usepackage[T1]{fontenc}    % use 8-bit T1 fonts
\usepackage{hyperref}       % hyperlinks
\usepackage{url}            % simple URL typesetting
\usepackage{booktabs}       % professional-quality tables
\usepackage{amsfonts}       % blackboard math symbols
\usepackage{nicefrac}       % compact symbols for 1/2, etc.
\usepackage{microtype}      % microtypography
\usepackage{cleveref}       % smart cross-referencing
\usepackage{lipsum}         % Can be removed after putting your text content
\usepackage{graphicx}
\usepackage{natbib}
\usepackage{doi}
\usepackage{authblk}
\usepackage{subcaption}

\title{The impact of feature selection and transformation on machine learning methods in determining the credit scoring}

\author[a]{Oğuz Koç}
\author[b]{Ömür Uğur}
\author[c]{A. Sevtap Kestel}
\affil[ ]{Institute of Applied Mathematics \\
	 Middle East Technical University \\
	06800 Ankara, Turkey
}
\affil[a]{oguzkoc@metu.edu.tr}
\affil[b]{ougur@metu.edu.tr}
\affil[c]{skestel@metu.edu.tr}

\renewcommand{\shorttitle}{\textit{arXiv} Template}

\begin{document}
\maketitle

\begin{abstract}
Banks utilize credit scoring as an important indicator for the financial strength and the eligibility for credit. Scoring models aim to assign statistical odds or probabilities for predicting if there is a risk of nonpayment in relation to many other factors which may be involved in. This paper aims to illustrate the beneficial use of the eight machine learning (ML) methods (Support Vector Machine, Gaussian Na\"ive Bayes, Decision Trees, Random Forest, XGBoost, K-Nearest Neighbors, Multi-layer Perceptron Neural Networks) and Logistic Regression in finding the default risk as well as the features contributing to it. An extensive comparison is made in three aspects: (i) which ML models with and without its own wrapper feature selection performs the best; (ii)  how feature selection combined with appropriate data scaling method influences the performance; (iii) which of the most successful combination (algorithm, feature selection and scaling) delivers the best validation indicators such as accuracy rate, Type I and II errors and AUC. An open-access credit scoring default risk data sets on German and Australian cases are taken into account for which we determine the best method, scaling and features contributing to default risk best and compare our findings with literature the ones in related. We illustrate the positive contribution of the selection method and scaling on the performance indicators compared to the existing literature.
\end{abstract}

% keywords can be removed

\section{Introduction}

Machine Learning (ML) methods offer a wide range of approaches to determine the characteristics, specifications, and subcategorizations of variables by taking into account the associations among each other ~\citep{mueller2016machine}. Prediction accuracy of these methods is sensitive to the selection of the features, methods, size of the data and transformation properties. Additionally, such variability in ML gives many alternatives to the users to increase the prediction power. Certain statistical tests are used to depict the accuracy of these methods ~\citep{han2011data}.

The implementation of ML becomes very attractive, especially after the development of high speed computers which allows to process huge number of data sets ~\citep{alpaydin2016machine}. As much as the other disciplines, financial sector benefits the use of ML methods as well as diversity in algorithms. A very important application area, is the banking sector whose financial risk is highly uncertain and difficult to predict due to its complex structure in the multi-dimensional factors involved in. Banks are prone to high default risk, mainly due to credit risk which may arise from market risk and economic recessions~\citep{lessmann2015benchmarking}. A recent history (subprime crisis in 2008) has shown that, as a result of the unrestrained growth of banking sector and financial globalization, the financial crisis arise from housing market fluctuations in the USA, and spread worldwide causing big financial disasters at every level in the world. Consequently, lack of risk management in banking sector before subprime crisis results in large losses. After this milestone in financial history, regulatory framework such as Basel II and then III are developed, and risk management becomes one of the main tool determining contestability of banks~\citep{gatzert2012comparative}. Besides, the current and future risk levels of the bank's loan portfolio are crucial in terms of its competitiveness, resilience to financial crises and profitability. These requirements urge banks to quantify the amount of risk they are prone to by determining the credit risk at institutional and individual bases. Banks can make credit classification in two different ways. The first is based solely on the expert judgment of the credit analyst. Another is the credit scoring which is basically the classification criteria  based on the current and historical financial strength of the client~\citep{crook1996credit}. Credit assessment through the risk management system should be robust, reliable and modernized with respect to the changes in data science~\citep{anderson2007credit}. For this reason, the investigation of the most suitable and current ML classification algorithms fulfilling these requirements gain importance during the last decades. 

In the implementation of ML methods, four aspects contribute to the success in prediction: (i) the ML method chosen, (ii) the feature selection, (iii) the data transformation, and (iv) the validation structure. It is apparent that the combination of these four aspects creates a huge amount of trials and varying computation times to come up with the best predicting method.  In literature, there exists vast amount of studies on measuring the credit risk, mostly taking into account, linear discriminant analysis, factor analysis, logistic regression, artificial intelligence such as expert systems and neural networks, Bayes classifier, nearest neighbor, and classification trees. It is found that neural networks and XGBoost (XGB) algorithms are more accurate in classification than other techniques, especially on  loan assessment and predicting the default probability \citep{xia2017boosted,marcano2011artificial,khashei2013bi,yeh2009comparisons,munkhdalai2019empirical}. The recent and recognizable studies comparing the most suitable ML methods based on real data sets point out that random forest (RF), Gradient Boosting (GB), Decision Tree (DT), Neural Network (NN), Support vector Machine (SVM), K-Nearest Neighborhood (KNN) and some of their variations yield relatively good results compared to the traditional methods such as logistic regression \citep{brown2012experimental,lessmann2015benchmarking,ince2009comparison,baesens2003benchmarking,marques2012exploring}. In feature selection area on ML algorithms, variety of methods, such as wrapper, filter and hybrid algorithms, are applied by considering performance indicators like the highest median and the lowest standard deviation \citep{han2011data,xue2015comprehensive,chen2010combination,suto2016comparison,nnamoko2014evaluation}. Especially, methods with Genetic Algorithms (GA) and Particle Swarm Optimization (PSO) and na\"ive Bayes (NB) are found to have the highest performance achievement against the widely used filter methods t-Test, linear discriminant analysis (LDA),  logistic regression (LR), information gain, and rough set theory based ones in terms of receivers operating curve (ROC) and accuracy ratio. As the third aspect in the ML analyses, the effect of data transformation is not considered extensively in the literature. The most well-known transformation methods used in literature are Standard and Min-Max transformation methods. Besides, Median Norm, Sigmoid Norm, Statistical Column Normalization are also implemented to see the influence of the data transformation on the variables to point out the effectiveness of transformation on the accuracy ~\citep{szymanski2004recursive,jayalakshmidr}). The ML algorithms require justification of goodness of fit through statistical tests (validation criteria) which are imposed on training data whose sensitivity changes with respect to its number of observations. The most common selection is 80-20 \% separation of train and test groups, which may not be effective for small samples~\citep{zheng2018feature,yu2010comparative}.

Having the motivation on figuring out the best ML algorithm together with the best combination of feature selection, data transformation and train-test separation in credit risk scoring, this paper aims to obtain more explanatory results in terms of dimentionality, and investigate capability of detection significant features when wrapper feature selection (WFS) method is used with ML algorithms. We also aim to analyse the contribution of hyper-parameter optimization of these parameters which are included in all ML algorithms and can be manually adjusted with Grid Search (GS) along with the four types of data transformation to these algorithms and methods. For illustration of our approach, the German and Australian credit data sets~\citep{Dua:2019} are studied due to their online accessibility and availability as well as to make comparison of our findings with the literature~\citep{liang2015effect, marques2012exploring, xia2017boosted, yu2010comparative, tsai2009feature, shen2019novel, biswas2018rule, ramasamy2018online}.

The findings of this paper guide the researchers both in inducing the important characteristics in credit scoring as well as the correct implementation of the factors to have the maximum accuracy in ML applications.   

The paper is organized as follows: Section 2 shortly describes metrics, feature selection and parameter optimization, whereas a brief definition of ML methods are introduced in Section 3. The definition of validation criteria, experimental design approaches along with the introduction to German and Australian data sets are presented in Section 4. Outcomes of the application of the ML algorithms are explained and illustrated in Section 5. There, we present the accuracy and efficiency indicators in terms of the significant characteristics in three aspects consisting of large variety of different approaches. The impact of data normalization techniques to the efficiency of the ML algorithm, the effect of wrapper methods on ML and the examination on how GS parameter optimization affects these algorithms are explained in details. The final section concludes the paper with a summary and remarks.

\section{Data Processing, Feature Selection and Parameter Optimization}

In preprocessing phase on the data set, feature scaling is a vital part that could be easily ignored by the analysts. DT, RF and XGB are some of tree based algorithms that do not require feature scaling. These algorithms are insensible against to scaling methods, whereas the most of ML and optimization methods work more efficiently when attributes are on the same range ~\citep{raschka2015python}. For instance, different attributes are typically measured on various ranges, and if there is a use of Euclidean distance or Manhattan distance metric, it can cause some features completely outweigh the others that have smaller range of measurement  ~\citep{witten2016data}. Normalization and standardization are two of the most common methods for bringing distinct attributes to the same scale  ~\citep{raschka2017python}. 

We consider natural log, Box-Cox, Min-Max and Standard scaling methods in our analyzes. Box-Cox and natural log are the members from the power transformation family. These two techniques are applied to make the data distribution bell shaped by shrinking the higher values and expanding it to the lower values. These are also useful for heavy tailed distributions ~\citep{zheng2018feature}. Min-Max scaling function provides a linear transformation of the initial data. It maintains the connection between the values of initial data ~\citep{han2011data}. Min-max scaling also squeezes (or stretches) all feature values to be within the range between 0 and 1 by considering the maximum value and minimum of an attribute column  ~\citep{zheng2018feature}. In standard scale feature columns are set in the midst at standard deviation one and mean zero. In contrast to Min-Max scaling, the standard scaling sustain useful outlier information and makes the algorithms more robust~\citep{raschka2015python}.

Feature selection is one of the most influential analysis for the performance of the model, since irrelevant or partially relevant features can cause negative effect. Besides, this analysis provides various benefits such as reduction in run time and obtaining more accurate results  ~\citep{stanczyk2015feature}. In our study, we use wrapper attribute selection approach (WFS) in which the learning algorithm is wrapped into variable subset selection procedure. It would be simple to make an independent evaluation of an attribute subset if there was a silver bullet way to determine when an attribute was related to the class. Yet, there is no widely accepted measure of significance, even though several variety of measurements have been proposed  ~\citep{zheng2018feature}. The general concept of the wrapper method is to pick a subset of features using a learning algorithm as a part of the evaluation function that can be used with any machine learning algorithm  ~\citep{bolon2016feature}. To prevent overfitting we apply this method with K-Fold Cross validation (KFCV). Also, application of this method is conducted by the Sequential Forward Selection (SFS). This begins with the induction of machine learning algorithms for \textit{n} single-attribute subsets from which the best performer is selected after the first selection phase. Among the $ n-1 $ attributes the most fitted one with the already selected feature is spotted, and this attribute is added to the subset. This process goes on until the number of subsets reaches to \textit{n}. In that way, whole attribute combinations are validated and the subset that has the highest score, such as accuracy or minimum weighted error is detected.

In ML, there are basically two kinds of parameter estimates. The first one is learned from the training data, as the weights of neurons in MLP; the other is the hyper-parameters of a learning algorithm that are optimized separately, such as the number of trees in RF ~\citep{raschka2015python}. The search of the possible combinations of hyper-parameters, and where to look for an optimum is a cumbersome task. An hyper-parameter space may contain value combinations that perform better or worse. Even after detecting a good combination, there may always be a better combination of the parameters due to being a local optima. A practical way to overcome this problem, verifying hyper-parameters for an algorithm applied to specific data, is to test them in large variety of KFCV trials, and to pick the best combination based on the highest accuracy rate. 

The Grid Search (GS) makes this process easier by allowing the user by sampling the range of possible values to input into the algorithm and to spot when the general error rate is in minimum ~\citep{mueller2016machine}. This is a brute-force approach in which a list of prespecified values belonging to different hyper-parameters is used, and the model performance is assessed based on the highest accuracy among the combination of those parameters. 

To obtain more robust and general results, GS is commonly applied with a 10-fold cross validation ~\citep{raschka2017python} approach. 

\section{Machine Learning Algorithms}

Mathematical and statistical algorithms of eight commonly used and employed in this paper are briefly summarized. These are, LR, Gaussian Na\"ive Bayes (GNB), SVM, DT, RF, XGB, KNN and Multi-layer Perceptron Neural Networks (MLP).\\

\textbf{Logistic Regression} (LR) is the most commonly used and back-dated method having the form

\begin{equation} P(D|X_1,X_2,\cdots,X_n)=\frac{e^{\beta_0 + \beta_1X_1+\cdots+\beta_nX_n}}{1+e^{\beta_0 + \beta_1X_1+\cdots+\beta_nX_n}}, \end{equation}
where $X_1,X_2,...,X_n$ are exogenous and $D$ is the binary dependent variable. Central point in this transformation is that it is designed to identify a probability that is always a number in the range between $0$ and $1$ due to the S-shape of the logistic function. Hence, with this design the function can generate a probability value. The logistic model, therefore, is set up to ensure that whatever calculation of risk is obtained, it is a number between $0$ and $1$ ~\citep{kleinbaum2002logistic}. The main drawback in LR is to justify certain statistical conditions for the validity of the model. However, based on the studies done in LR as a ML method these diagnostic tests are mostly neglected.

\textbf{Gaussian Na\"ive Bayes} (GNB) is a statistical classifier that has a white box nature which means the ML process is transparent and clear understanding of how they behave ~\citep{li2012overview}. The posterior probability of every samples is calculated by considering their particular classes ~\citep{friedman1997bayesian} which assign the most probable class to a specified instance defined by its attributes. The algorithm is called Na\"ive since it assumes that features are independent for the corresponding class ~\citep{rish2001empirical}. The conditional probability of an observation belong to class $C$, $C$ is a subset of $E$ = $(x_1,x_2,...,x_n)$, is given as  
\begin{equation}P(C|E) = \frac{{P(E|C)}{P(C)}}{P(E)}. \end{equation}
Here, in practice, the conditional distribution is assumed to be normally distributed to handle the continuous variables at which the Gaussian transformation is required as follows:

\begin{equation}\label{eq:5.2}P(E|c) = \frac{1}{\sqrt{2\pi\sigma_c^2}}{e^{-\frac{(x-\mu_c)^2}{2\sigma_c^2}}}\end{equation}
Here, $\mu_c$ and $\sigma_c$ are the mean and standard deviation of of class $c$, respectively. With this transformation the conditional probability numeric attributes can be calculated.

\textbf{Support Vector Machine} (SVM) converts the initial training data into a higher dimension alone and employs a nonlinear mapping. Within this new dimension, it searches for the linear optimal separating hyperplane (i.e., a decision boundary separating the tuples of one class from others.). In this high dimensional space, the algorithm search decision boundary which separates linearly the variables belonging to different classes. The SVM decides the hyperplane location and margins by using the support vectors from the training points ~\citep{han2011data}. A dividing hyperplane can be identified as 
\begin{equation}\label{eq:5.23}W\cdot X+b=0,\end{equation}
where $W = (w_1, w_2, \cdots, w_n)$ is a weight vector, $n$ is the number of features; and $b$ is a scalar, often referred to as a bias. If any point lies above the dividing hyperplane (Eq:\ref{eq:5.23}), these points satisfy the following condition when $b=w_0$

\begin{equation}\label{eq:5.25} w_0+\sum w_ix_i>0.\end{equation}

\textbf{Decision Tree} (DT) consists of internal nodes, branches and leaf which seems like a tree structure. Each internal node indicates a test on a feature, all leaves symbolize a class label, and each branch defines an outcome. In top of the tree structure, there is a root node at which the splitting is started~\citep{han2011data}. DT's are structured general to specific information by the usage of training data. For instance, in the case that all observations within a node belonging to same class, requires no splitting, because when a random dividing is occurred, it does not decrease the impurity ratio measured by Gini or Entropy. Hence, the node becomes a leaf and then the branching is completed ~\citep{apte1997data}.

The description of \textit{info}$(S)$ reflects the possible information occurred by splitting $t$ observation into $n$ sub nodes. For Entropy, it is defined as

\begin{equation} \label{eq:5.53} \textit{info} \ (S)=-\sum_{i=1}^n \frac{|t_i|}{|t|} \times \log_2 \left( \frac{|t_i|}{|t|} \right).\end{equation}

\textbf{Random Forest} (RF) allows a large number of classification trees that are growing based on Breiman's CART tree~\citep{breiman2001random} approach which requires placing the input variable down each of the trees in the forest to classify newly added test object from an input variable. Every tree in the forest gives a classification vote to classify an instance to appropriate category. The forest selects the class that the very instance belongs to by considering the most votes over all the trees. In training process, the amount of instances in the training set is sampled randomly by replacement from the initial dataset, called "bootstrap". In each step, the training set is sampled to build a new tree~\citep{breiman2015random}. When a bag of $B$ trees, $\hat{f}^{b}(x)$, is constructed, and then these trees are used for determination of an observation's class with the ultimate function, $\hat{f}_{bag}(x)$, as averaging overall ($B$) trees

\begin{equation}\hat{f}_{bag}(x)=\frac{1}{B}\sum_{b=1}^B\hat{f}^{b}(x).\end{equation}.

%\begin{aligned}
%\hat{y_i}^1&=f_1(x_1)=\hat{y_i}^0+f_1(x_i)\\
%\hat{y_i}^2&=f_1(x_i)+f_2(x_i)=\hat{y_i}^1+f_2(x_i)\\
%&\vdots\\
%\end{aligned}

\textbf{XGBoost} (XGB) is an improved version of the Gradient Boosting algorithm based on CART decision tree approach. In this method, regularized objective and sparsity-awareness are the two novel approaches. This algorithm can also handle missing inputs, frequently zero values and binary data with default direction approach. The other powerful sides of this algorithm are: it is faster than other most popular machine learning methods and prevents overfitting by shrinkage and column subsampling approaches ~\citep{chen2016xgboost}. The model is trained in an additive process: let $\hat{y}_i^{(t)}$ be the predicted value of the $i$-th sample at the $t$-th iteration as
\begin{equation}\hat{y}_i^{(t)}=\sum_{k=1}^K f_k(x_i)=\hat{y_i}^{(t-1)}+f_t(x_i)\end{equation}
where $\hat{y}_i^{(0)}=0$. By adding $f_t(x_i)=\hat{y}_i^{(t)}-\hat{y}_i^{(t-1)}$ step by step to improve this ensemble model, the algorithm minimizes the following:
\begin{equation}\mathcal{L}^{(t)}=\sum_{t=1}^nl(y_i,\hat{y}_i^{(t-1)}+f_t(x_i))+\Omega(f_t)\end{equation}
Here,  $\Omega$ penalizes the complexity of the CART decision tree functions and it is defined as 
\begin{equation}\label{eq:5.10}\Omega(f)=\gamma T+\frac{1}{2}\lambda ||w||^2.\end{equation}
To prevent over-fitting conclusive learned continuous scores ($w_i$) for every tree $f_t(x)$, $\lambda$ parameter is used for smoothing, and the complexity of each trees is reduced via $\gamma$.

\textbf{K-Nearest Neighbor} (KNN) is a memory based method and requires no model to be fitted. In other words, KNN does not make any assumption on data to produce a model. Given a testing point, $x_{ij}$, belongs to $i^{th}$ observation's $j^{th}$ feature in the sample space, we can find the closest $k$ points, $x_{ij}, i=1,\cdots,k$, based on a distance function used. Than we assign the instance ($x_{ij}$) to suitable class by usage of majority vote over all the $k$ neighbors ~\citep{trevor2009elements}. The closeness between two tuples or points is described in terms of a distance metric, such as Euclidean distance,
\begin{equation}\label{eq:5.76}dist(X_1,X_2)= \sqrt{\sum_{i=1}^n({x_{1i}-x_{2i}})^2},\end{equation}
where $X_1=(x_{11},x_{12},...,x_{1n})$ and $X_2= (x_{21},x_{22},...,x_{2n})$.

\begin{table*}[h]
	\caption{The variables in German and Australian credit data with their features}
	\label{tab:abc}
	\counterwithout{table}{section}
	\centering
	\scalebox{1}{
		\begin{tabular}{lllllll}
			\hline
			& \textbf{German Data}&\multicolumn{1}{c|}{} & &  \textbf{Australian Data} & \\
			\hline
			\textbf{Attribute} &  \textbf{Description}  &  \multicolumn{1}{c|}{\textbf{Type}}& & \textbf{Attribute} &  \textbf{Type} \\ 
			\hline
			1                   &  Age (years)                &  \multicolumn{1}{c|}{Continuous} & & 1 (A) & Categorical (0,1)\\
			2                   &  Duration (month)           &  \multicolumn{1}{c|}{Continuous} & & 2 (B) & Continuous \\
			3                   &  Credit history              &  \multicolumn{1}{c|}{Categorical} & & 3 (C) & Continuous\\
			4                   &  Purpose                     &  \multicolumn{1}{c|}{Categorical} & & 4 (D) & Categorical (1-3)\\
			5                   &  Credit amount               &  \multicolumn{1}{c|}{Continuous} & & 5 (E) & Categorical (1-14)\\
			6                   &  Savings account/bonds       &  \multicolumn{1}{c|}{Categorical} & & 6 (F) & Categorical (1-9)\\
			7                   &  Present employment since    &  \multicolumn{1}{c|}{Categorical} & & 7 (G) & Continuous\\
			8                   &  Instalment rate             &  \multicolumn{1}{c|}{Continuous} & & 8 (H) & Continuous [0,1]\\
			9                   &  Personal status and sex     &  \multicolumn{1}{c|}{Categorical} & & 9 (I) & Continuous [0,1]\\
			10                  &   Other debtors/guarantors    &  \multicolumn{1}{c|}{Categorical} & & 10 (J) & Continuous\\
			11                  &  Present residence since     &  \multicolumn{1}{c|}{Continuous} & & 11 (K) & Continuous [0,1]\\
			12                  &  Property                    &  \multicolumn{1}{c|}{Categorical} & & 12 (L) & Continuous [1-3]\\
			13                  &  Status of existing checking account &  \multicolumn{1}{c|}{Categorical} & & 13 (M) & Continuous&\\
			14                  &  Telephone                     &  \multicolumn{1}{c|}{Categorical} & & 14 (N) & Continuous &\\
			15                  &  Other instalment plans      &  \multicolumn{1}{c|}{Categorical} & & 15 Default State &  Categorical (0,1) \\
			16                  &  Housing                     &  \multicolumn{1}{c|}{Continuous} & & & \\
			17                  &  Number of existing credits at this bank &  \multicolumn{1}{c|}{Categorical} & & & \\
			18                  &  Foreign worker                &  \multicolumn{1}{c|}{Continuous} & & & \\
			19                  &  Job                         &  \multicolumn{1}{c|}{Categorical} & & & \\
			20                  &  Number of dependants        &  \multicolumn{1}{c|}{Categorical} & & & \\
			21                  &  Class labels        				 &  \multicolumn{1}{c|}{Categorical} & & & \\ 
			22					&  Binary Default State        &  \multicolumn{1}{c|}{Categorical} & & & &\\ \hline
	\end{tabular}}
\end{table*}

\textbf{Multi-layer Perceptron Neural Networks} (MLP) is an extended version of feed forward Neural networks. This framework ordinarily involves an input layer, one or more hidden layers, and the output layer. Initiation of the data samples takes place in the input layer. This layer feeds the samples to hidden layer without any change. After that in hidden layer, every node receives the outputs from the previous layer, make some assumptions, and deliver the results to the next layer. Finally, output layer generates the last outputs that end up with a classification or regression ~\citep{grus2019data}. Neurons that is not input neuron have weights and a bias associated to every one of its inputs . In order to give more straightforward definition of the structure, the bias vectors are added to vector weights and a bias that is always equal to one is used for every neuron. The outcomes of the neurons' inputs and weights will be summed up. The weight of the link from the $k$-th neuron in the $(l-1)$-st layer to the $j$-th neuron in the $j$-th layer is defined by $w_{jk}^l$ and $b_j^l$ is the bias and $a_j^l$ is the activation of the $j$-th neuron in the $l$-th layer. The $a_j^l$ activations of the $j$-th neuron in the $l$-th layer are linked to the $(l-1)$-th layer activations by
\begin{equation}a_l^l=\sigma \left( \sum_kw_{jk}^la_k^{l-1}+b_j^l \right).\end{equation}
Here, the summation of all $k$ neurons in the $(l-1)$-th layer is given, and $\sigma$ is the sigmoid activation function. The weight matrix inputs, $w^l$, are the weights that are connected to the $l$-st layer of neurons, and a bias vector will be identified as, $b^l$~\citep{nielsen2015neural}. The weighted input to the neurons in layer $l$ is defined as 
\begin{equation}z^l=w^la^{l-1}+b^l.\end{equation}
The back propagation algorithm is employed to adjust the weights by using a cost function in the training stage:
\begin{equation}\label{eq:5.81} C=\frac{1}{2n}\sum_{x}^{} {\left\|y(x) -a^L(x)\right\|}^2,\end{equation}
where $L$ is the number of layers, $a^L(x)$ is the vector of activation function outputs and $y(x)$ is the vector of target values for the $n$ observation in the training set.

\section{Validation Methods and Credit Scoring Data}

Validation criteria are the measures for assessing how good or how accurate a classifier is at predicting the class labels of instances. The Accuracy Ratio (AR), Type I and II error rates and Area Under Curve (AUC) ratio are used. Sklearn-learn libraries~\citep{scikit-learn} are used to conduct the experiments for eight machine learning algorithms.

Ratio of the number of truly predicted observations and the total number of observation gives the AR. Type I error rate can be calculated by dividing the number of incorrectly classified observations that belong to Class I by total number of observations in Class I. Type II error is defined as the proportion of number of falsely detected Class 0 to total number of cases in Class 0.  Additionally, ROC curve for a given model shows the trade-off between the true positive rate (TPR) and the false positive rate (FPR). The area under the ROC curve is a measure of the accuracy of the model which is called as AUC ratio which is most commonly known measure in the literature ~\citep{han2011data}.

On the other hand, in K-fold cross-validation, the present samples are split at random into $k$ disassociated subsets,\newline $D_1, D_2, \cdots, D_k$, where each has roughly equal number of sample. In this validation process, testing and training is executed $k$ times. In step $i$, split part $D_i$ is kept as a validation set, and the rest of the data samples are used to train the ML model.

An important decision in the analyses requires the experiments to be separated into two sets as Train and Test parts. The proportion of the separation plays an important role. The most of the studies in literature considers as randomly divided subset of data set having 80\% as the training and 20\% as the test sets. This selection criteria is crucial depending on the size of the data and the frequency of ones and zeros (Class 1 and Class 0) in the overall set. Literature displays also different partition among train and test data. The study ~\citep{yu2010comparative} shows partition of data set starting from 80\% at each incremental percent to 90\% has influence on the accuracy indicators.

In our analyzes, we aim to show the classification ability of the eight ML algorithms, and impact of the four data transformation techniques on these algorithms. Beside, we employ three approaches in our analysis. The first one is the application of GS for parameter optimization, the second one is the implementation of Wrapper Feature Selection (WFS) method for feature selection, and the last one is the jointly use of WFS and GS. 10-fold cross validation is applied to avoid over fitting in these approaches. Each of the experiments are validated on three randomly divided train and test subsets. The data sets are randomly divided into $90\%$ as the train and $10\%$ as the tests set by maintaining the distribution of classes same as the original data set. In this way, we can prevent the insufficiency in terms of default or non-default samples on train or test data sets. Average of AUC, AR, Type I and Type II errors of these three cases is used to compare the effectiveness of the algorithms and methods.

\subsection{German and Australian Credit Data}

Due to the regulations and laws on the protection of personal data in banking sector and  institutional confidentiality reasons, it is very difficult to receive credit history from any financial institutions. For this reason, two real-world data sets, German and Australian, which are provided by the UCI machine learning data repository~\citep{Dua:2019}, are used to conduct our proposed study. Since these data sets are frequently used in the literature, it also sets a good base to make comparison to other published works. 

\textbf{German Dataset} contains 24 attributes and 1000 observations 700 of which belong to Class 1 (No default) and 300 of the observations to Class 0 (Default). In our study we use the one adjusted to be suitable for working with  also algorithms that can work only numerical variables (not categorical ones). Seven of these attributes are numerical where as the others are categorical variables. The definition of the attributes are given and details are listed in the left of Table ~\ref{tab:abc}. However, \textbf{Australian Dataset} consists of credit card applications of 690 observations, 307 to be classified as Class 0 and the rest as Class 1. The data source and the attributes contained within are labeled by letter codes. There are six numeric and 8 categorical attributes that are also appropriate for the algorithms (right panel in Table ~\ref{tab:abc}) whose details are covered or transformed to protect personal information, remain only within the type of the variables.

\section{Application}

In this part, we investigate which of the eight classification methods is superior to others in terms of consumer credit classification. We also examine the effect of GS and WFS methods on the performance of these classification algorithms, and observe how four different data conversion techniques influence the effectiveness of these algorithms. WFS is applied to each algorithm with use of sequential forward selection (SFS) method. Experimental design is made on the separation of 90\% for training and 10\% for testing by maintaining the proportion of each classes in both sets. Data transformations are implemented before splitting the sets for training. In order to compare the results, data sets are divided according to certain random states. For each case, three different train and test sets are generated according to these random states, and average of these three test results' are considered for comparison in terms of AUC, AR, Type I and Type II error rates. The summary of the analyses is listed in Tables ~\ref{tab:ConMat_D} and ~\ref{tab:ConMat_AU} for German and Australian data, respectively. These tables indicate the ML algorithms with different combinations of data transformations (original-no scaling, standardized, Min-Max, Box-Cox and natural log), the ML methods (GNB, SVM, MLP, DT, RF, XGB, LR, KNN) and feature selections (WSF, GS, WSF and GS jointly) in terms of their efficiency rates in percents. Illustrative comparisons of the findings are also plotted and presented in Figure \ref{fig:mytable}. These figures depict the accuracy of each method used with respect to the data transformations, hyper parameter optimization and feature selections.

In terms of data transformations, it is seen that impact of these techniques vary according to data set and ML algorithms. The interpretations of these findings are discussed with respect to (i) performance of ML method, (ii) influence of transformations, (iii) impact of feature selection method, (iv) the computation time and (v) performance in case of highly unbalanced class distribution. Below, we emphasise on the important analysis. The details of these analyses can also be found in ~\citep{Koc2019}.

\begin{table*}[h]
	\centering
	\caption{The efficiency indicators of ML methods for \textbf{German} data (in \%)}
	\counterwithout{table}{section}
	\label{tab:ConMat_D}
	\resizebox{\textwidth}{!}{%
		\begin{tabular}{lcccccccccccccc}
			\hline
			\textbf{Algorithm} & \multicolumn{2}{c}{\textbf{No Scale}} & \multicolumn{2}{c}{\textbf{Standard}} & \multicolumn{2}{c}{\textbf{Min-Max}} & \multicolumn{2}{c}{\textbf{Box-Cox}} & \multicolumn{2}{c}{\textbf{natural log}} & \multicolumn{2}{c}{\textbf{Min-Max (all)}} & \multicolumn{2}{c}{\textbf{Standard (all)}} \\ \hline
			\textbf{Classifier} & \multicolumn{1}{l}{\textbf{Accuracy}} & \multicolumn{1}{l}{\textbf{Type I}} & \multicolumn{1}{l}{\textbf{Accuracy}} & \multicolumn{1}{l}{\textbf{Type I}} & \multicolumn{1}{l}{\textbf{Accuracy}} & \multicolumn{1}{l}{\textbf{Type I}} & \multicolumn{1}{l}{\textbf{Accuracy}} & \multicolumn{1}{l}{\textbf{Type I}} & \multicolumn{1}{l}{\textbf{Accuracy}} & \multicolumn{1}{l}{\textbf{Type I}} & \multicolumn{1}{l}{\textbf{Accuracy}} & \multicolumn{1}{l}{\textbf{Type I}} & \multicolumn{1}{l}{\textbf{Accuracy}} & \multicolumn{1}{l}{\textbf{Type I}} \\ \hline
			GNB & 75.33 & 38.89 & 75.33 & 38.89 & 75.33 & 38.89 & 72.33 & 38.89 & 73.33 & 37.78 & 75.33 & 38.89 & 75.33 & 38.89 \\
			GNB+GS & 76.67 & 52.22 & 77.33 & 40.00 & 75.33 & 44.44 & 74.33 & 43.33 & 73.33 & 42.22 & 75.00 & 42.22 & 75.33 & 38.89 \\
			GNB+Wr & 76.00 & 46.67 & 76.00 & 46.67 & 76.00 & 46.67 & 77.33 & 43.33 & 75.33 & 50.00 & 76.00 & 46.67 & 76.00 & 46.67 \\
			GNB+GS\&Wr & 76.00 & 46.67 & 76.00 & 46.67 & 76.00 & 46.67 & 77.33 & 43.33 & 75.33 & 50.00 & 76.00 & 46.67 & 76.00 & 46.67 \\ \hline
			SVM & 71.00 & 92.22 & 76.00 & 60.00 & 74.67 & 66.67 & 75.67 & 60.00 & 76.00 & 61.11 & 73.67 & 74.44 & 76.67 & 58.89 \\
			SVM+GS & 79.33 & 50.00 & 76.67 & 57.78 & 78.00 & 50.00 & 78.00 & 50.00 & 78.67 & 50.00 & 76.33 & 55.56 & 77.00 & 55.56 \\
			SVM+Wr & 72.67 & 71.11 & 77.00 & 57.78 & 71.67 & 73.33 & 74.33 & 62.22 & 75.67 & 58.89 & 74.33 & 68.89 & 76.67 & 61.11 \\
			SVM+GS\&Wr & 72.00 & 71.11 & 77.00 & 56.67 & 72.67 & 68.89 & 75.33 & 60.00 & 78.00 & 56.67 & 75.00 & 67.78 & 77.33 & 57.78 \\ \hline
			MLP & 78.67 & 48.89 & 81.00 & 41.11 & 77.00 & 54.44 & 79.33 & 46.67 & 78.33 & 47.78 & 77.33 & 50.00 & 78.00 & 41.11 \\
			MLP+GS & 81.00 & 46.67 & 82.00 & 40.00 & 74.33 & 61.11 & 77.33 & 51.11 & 77.33 & 51.11 & 78.33 & 50.00 & 77.67 & 52.22 \\
			MLP+Wr & 79.67 & 48.89 & 79.33 & 43.33 & 75.33 & 54.44 & 77.33 & 50.00 & 75.67 & 54.44 & 76.00 & 53.33 & 79.00 & 47.78 \\
			MLP+GS\&Wr & 80.00 & 47.78 & 79.67 & 42.22 & 74.33 & 56.67 & 76.33 & 53.33 & 75.33 & 56.67 & 76.33 & 52.22 & 79.33 & 47.78 \\ \hline
			DT & 66.33 & 57.78 & 66.33 & 57.78 & 66.33 & 57.78 & 66.33 & 57.78 & 66.33 & 57.78 & 66.33 & 57.78 & 66.33 & 57.78 \\
			DT+GS & 77.00 & 56.67 & 77.00 & 56.67 & 76.67 & 55.56 & 76.67 & 55.56 & 76.67 & 55.56 & 76.67 & 55.56 & 76.67 & 55.56 \\
			DT+Wr & 72.67 & 67.78 & 72.67 & 67.78 & 72.67 & 67.78 & 72.67 & 67.78 & 72.67 & 67.78 & 72.67 & 67.78 & 72.67 & 67.78 \\
			DT+GS\&Wr & 73.33 & 67.78 & 73.33 & 67.78 & 73.33 & 67.78 & 73.33 & 67.78 & 73.33 & 67.78 & 73.33 & 67.78 & 73.33 & 67.78 \\ \hline
			RF & 74.67 & 51.11 & 75.00 & 51.11 & 74.33 & 48.89 & 74.67 & 50.00 & 75.00 & 51.11 & 73.60 & 51.11 & 75.00 & 51.11 \\
			RF+GS & 79.00 & 55.56 & 79.00 & 53.33 & 78.67 & 55.56 & 80.33 & 50.00 & 80.67 & 50.00 & 80.33 & 50.00 & 79.00 & 53.33 \\
			RF+Wr & 75.33 & 44.44 & 73.33 & 48.89 & 75.00 & 52.22 & 74.33 & 48.89 & 74.33 & 44.44 & 72.00 & 48.89 & 74.00 & 44.44 \\
			RF+GS\&Wr & 76.33 & 55.56 & 77.67 & 54.44 & 77.67 & 54.44 & 74.00 & 63.33 & 75.33 & 61.11 & 77.67 & 54.44 & 76.33 & 55.56 \\ \hline
			XGB & 79.67 & 51.11 & 79.67 & 51.11 & 79.67 & 51.11 & 79.67 & 51.11 & 79.67 & 51.11 & 79.67 & 51.11 & 79.67 & 51.11 \\
			XGB+GS & 81.67 & 45.56 & 81.67 & 45.56 & 81.67 & 45.56 & 81.67 & 45.56 & 81.67 & 45.56 & 81.67 & 45.56 & 81.67 & 45.56 \\
			XGB+Wr & 78.67 & 51.11 & 78.67 & 51.11 & 78.67 & 51.11 & 78.67 & 51.11 & 78.67 & 51.11 & 78.67 & 51.11 & 78.67 & 51.11 \\
			XGB+GS\&Wr & 78.33 & 53.33 & 78.33 & 53.33 & 78.33 & 53.33 & 78.33 & 53.33 & 78.33 & 53.33 & 78.33 & 53.33 & 78.33 & 53.33 \\ \hline
			LR & 74.67 & 58.89 & 77.33 & 55.56 & 77.67 & 53.33 & 77.67 & 52.22 & 78.00 & 51.11 & 76.67 & 56.67 & 78.33 & 52.22 \\
			LR+GS & 78.00 & 53.33 & 78.00 & 53.33 & 78.33 & 52.22 & 78.00 & 52.22 & 78.00 & 51.11 & 78.33 & 52.22 & 77.67 & 54.44 \\
			LR+Wr & 77.00 & 54.44 & 78.00 & 51.11 & 77.00 & 56.67 & 76.67 & 53.33 & 75.33 & 56.67 & 76.33 & 55.56 & 78.00 & 52.22 \\
			LR+GS\&Wr & 76.67 & 55.56 & 78.00 & 51.11 & 77.33 & 55.56 & 76.33 & 54.44 & 75.33 & 56.67 & 76.33 & 55.56 & 78.00 & 52.22 \\ \hline
			KNN & 71.33 & 61.11 & 73.00 & 62.22 & 74.33 & 57.78 & 74.67 & 60.00 & 74.00 & 61.11 & 69.67 & 62.22 & 72.33 & 68.89 \\
			KNN+GS & 73.00 & 65.56 & 78.33 & 55.56 & 76.33 & 55.56 & 74.33 & 60.00 & 74.33 & 58.89 & 76.33 & 60.00 & 71.00 & 71.11 \\
			KNN+Wr & 73.33 & 60.00 & 74.67 & 53.33 & 73.33 & 57.78 & 74.00 & 54.44 & 72.67 & 60.00 & 72.00 & 56.67 & 74.33 & 56.67 \\
			KNN+GS\&Wr & 72.33 & 61.11 & 73.67 & 54.44 & 73.67 & 60.00 & 74.33 & 57.78 & 76.00 & 55.56 & 74.33 & 56.67 & 71.00 & 63.33 \\ \hline
		\end{tabular}%
}\end{table*}

\begin{table*}[h]
	\centering
	\caption{The efficiency indicators of ML methods for \textbf{Australian} data (in \%)}
	\counterwithout{table}{section}
	\label{tab:ConMat_AU}
	\resizebox{\textwidth}{!}{%
		\begin{tabular}{lcccccccccccccc}
			\hline
			\textbf{Algorithm} & \multicolumn{2}{c}{\textbf{No Scale}} & \multicolumn{2}{c}{\textbf{Standard}} & \multicolumn{2}{c}{\textbf{Min-Max}} & \multicolumn{2}{c}{\textbf{Box-Cox}} & \multicolumn{2}{c}{\textbf{natural log}} & \multicolumn{2}{c}{\textbf{Min-Max (all)}} & \multicolumn{2}{c}{\textbf{Standard (all)}} \\ \hline
			\textbf{Classifier} & \multicolumn{1}{l}{\textbf{Accuracy}} & \multicolumn{1}{l}{\textbf{Type I}} & \multicolumn{1}{l}{\textbf{Accuracy}} & \multicolumn{1}{l}{\textbf{Type I}} & \multicolumn{1}{l}{\textbf{Accuracy}} & \multicolumn{1}{l}{\textbf{Type I}} & \multicolumn{1}{l}{\textbf{Accuracy}} & \multicolumn{1}{l}{\textbf{Type I}} & \multicolumn{1}{l}{\textbf{Accuracy}} & \multicolumn{1}{l}{\textbf{Type I}} & \multicolumn{1}{l}{\textbf{Accuracy}} & \multicolumn{1}{l}{\textbf{Type I}} & \multicolumn{1}{l}{\textbf{Accuracy}} & \multicolumn{1}{l}{\textbf{Type I}} \\ \hline
			GNB & 78.57 & 8.55 & 80.95 & 8.55 & 80.95 & 8.55 & 85.71 & 10.26 & 85.71 & 10.26 & 80.95 & 8.55 & 80.95 & 8.55 \\
			GNB+GS & 80.48 & 8.55 & 81.43 & 7.69 & 87.14 & 12.82 & 85.71 & 10.26 & 85.71 & 10.26 & 88.10 & 17.09 & 82.38 & 7.69 \\
			GNB+Wr & 88.10 & 17.09 & 88.10 & 17.09 & 88.10 & 17.09 & 88.10 & 15.38 & 88.10 & 17.09 & 88.10 & 17.09 & 88.10 & 17.09 \\
			GNB+GS\&Wr & 88.10 & 17.09 & 88.57 & 16.24 & 88.10 & 17.09 & 88.10 & 15.38 & 88.10 & 16.24 & 88.10 & 17.09 & 88.10 & 17.09 \\ \hline
			SVM & 54.29 & 4.27 & 83.81 & 18.80 & 85.71 & 18.80 & 75.71 & 23.93 & 73.33 & 15.38 & 87.14 & 20.51 & 85.71 & 17.95 \\
			SVM+GS & 72.86 & 16.24 & 84.29 & 17.09 & 87.62 & 17.95 & 79.52 & 15.38 & 76.67 & 17.09 & 85.71 & 17.09 & 83.33 & 17.09 \\
			SVM+Wr & 85.71 & 18.80 & 84.76 & 17.95 & 86.67 & 19.66 & 84.76 & 19.66 & 86.19 & 18.80 & 86.67 & 20.51 & 88.10 & 17.95 \\
			SVM+GS\&Wr & 85.71 & 19.66 & 84.76 & 20.51 & 87.14 & 20.51 & 85.71 & 18.80 & 84.76 & 17.09 & 87.14 & 20.51 & 85.71 & 17.09 \\ \hline
			MLP & 74.29 & 24.79 & 85.24 & 16.24 & 84.29 & 17.95 & 86.67 & 19.66 & 86.67 & 17.09 & 86.19 & 16.24 & 85.71 & 14.53 \\
			MLP+GS &  79.05 & 17.95 & 85.71 & 16.24 & 87.14 & 18.80 & 87.62 & 14.53 & 85.71 & 13.68 & 86.19 & 17.09 & 86.19 & 13.68 \\
			MLP+Wr & 84.76 & 18.80 & 85.24 & 17.09 & 86.19 & 18.80 & 87.62 & 15.38 & 81.43 & 12.82 & 86.19 & 17.95 & 87.14 & 13.68 \\
			MLP+GS\&Wr & 85.24 & 16.24 & 85.24 & 16.24 & 86.19 & 18.80 & 85.71 & 16.24 & 84.29 & 17.09 & 86.67 & 17.95 & 87.62 & 12.82 \\ \hline
			DT & 86.19 & 13.68 & 86.19 & 13.68 & 86.19 & 13.68 & 86.19 & 13.68 & 86.19 & 13.68 & 86.19 & 13.68 & 86.19 & 13.68 \\
			DT+GS & 87.14 & 20.51 & 87.14 & 20.51 & 87.14 & 20.51 & 87.14 & 20.51 & 87.14 & 20.51 & 87.14 & 20.51 & 87.14 & 20.51 \\
			DT+Wr & 87.62 & 16.24 & 87.14 & 16.24 & 87.62 & 16.24 & 87.62 & 16.24 & 87.62 & 16.24 & 87.62 & 16.24 & 87.62 & 16.24 \\
			DT+GS\&Wr & 87.14 & 20.51 & 87.14 & 20.51 & 87.14 & 20.51 & 87.14 & 20.51 & 87.14 & 20.51 & 87.14 & 20.51 & 87.14 & 20.51 \\ \hline
			RF & 83.81 & 14.53 & 83.81 & 14.53 & 83.81 & 14.53 & 83.81 & 14.53 & 83.81 & 14.53 & 81.43 & 14.53 & 81.43 & 14.53 \\
			RF+GS & 86.67 & 12.82 & 86.67 & 12.82 & 86.67 & 12.82 & 86.19 & 13.68 & 86.19 & 13.68 & 86.67 & 12.82 & 86.67 & 12.82 \\
			RF+Wr & 85.71 & 14.53 & 85.24 & 14.93 & 85.24 & 14.93 & 86.19 & 13.68 & 85.24 & 14.93 & 85.24 & 14.93 & 85.24 & 14.93 \\
			RF+GS\&Wr & 85.71 & 14.53 & 86.19 & 13.68 & 86.67 & 12.82 & 85.71 & 14.53 & 86.67 & 12.82 & 87.14 & 13.68 & 87.14 & 13.68 \\ \hline
			XGB & 85.71 & 15.38 & 85.71 & 15.38 & 85.71 & 15.38 & 85.71 & 15.38 & 85.71 & 15.38 & 85.71 & 15.38 & 85.71 & 15.38 \\
			XGB+GS & 89.05 & 11.97 & 89.05 & 11.97 & 89.05 & 11.97 & 89.05 & 11.97 & 89.05 & 11.97 & 89.05 & 11.97 & 89.05 & 11.97 \\
			XGB+Wr & 87.14 & 16.24 & 87.14 & 16.24 & 87.14 & 16.24 & 87.14 & 16.24 & 87.14 & 16.24 & 87.14 & 16.24 & 87.14 & 16.24 \\
			XGB+GS\&Wr & 87.62 & 15.38 & 87.62 & 15.38 & 87.62 & 15.38 & 87.62 & 15.38 & 87.62 & 15.38 & 87.62 & 15.38 & 87.62 & 15.38 \\ \hline
			LR & 78.10 & 13.68 & 87.14 & 14.53 & 86.19 & 17.95 & 86.19 & 17.09 & 86.67 & 16.24 & 86.19 & 18.80 & 85.71 & 17.95 \\
			LR+GS & 86.19 & 14.53 & 87.62 & 14.53 & 86.19 & 17.95 & 86.67 & 16.24 & 87.14 & 15.38 & 86.19 & 18.80 & 85.24 & 17.95 \\
			LR+Wr &  86.67 & 17.95 & 87.62 & 16.24 & 86.19 & 17.95 & 86.67 & 17.95 & 86.19 & 17.09 & 87.14 & 17.95 & 87.62 & 16.24 \\
			LR+GS\&Wr & 86.67 & 17.95 & 87.62 & 16.24 & 85.71 & 17.95 & 86.67 & 17.95 & 86.19 & 17.09 & 87.14 & 17.95 & 88.10 & 15.38 \\ \hline
			KNN & 72.86 & 16.24 & 77.14 & 21.37 & 80.95 & 20.51 & 70.00 & 16.24 & 70.00 & 16.24 & 84.76 & 17.95 & 84.29 & 15.38 \\
			KNN+GS & 76.19 & 8.55 & 80.00 & 14.53 & 82.86 & 20.51 & 73.33 & 10.26 & 74.76 & 7.69 & 85.71 & 15.38 & 84.76 & 13.68 \\
			KNN+Wr & 80.95 & 15.38 & 85.24 & 11.11 & 89.05 & 12.82 & 83.81 & 14.53 & 83.33 & 17.09 & 87.62 & 14.53 & 86.19 & 15.38 \\
			KNN+GS\&Wr & 84.29 & 15.38 & 85.24 & 12.82 & 89.05 & 11.97 & 85.71 & 13.68 & 86.19 & 14.53 & 87.62 & 9.68 & 85.71 & 14.53 \\ \hline
		\end{tabular}%
	}
\end{table*}

\paragraph{Performance of ML algorithms} The performance of SVM is considerably improved with GS on both data sets compared to other states (79.33\% and  87.62\% of accuracy) (Figure~\ref{fig:SVM}). Although the WFS method leads to consistent results on Australian data, the same does not apply to German data. While KNN is more compatible with WFS on Australian data up to (89.05\% of accuracy), use of GS leads to better results in German one  (Figure~\ref{fig:KNN}). MLP gives higher accuracy rates with GS on both data sets (82.00\% and 87.62\% of accuracy) (Figure~\ref{fig:MLP}). WFS has a positive impact to MLP on Australian data, but the same is not observed for German data. Studies with LR support also that this algorithm gives better results with GS on both data (Figure~\ref{fig:LR}) whereas WFS\&GS is more suitable for Australian data (88.10\% of accuracy) than the other. GNB and GS combination gives a good classification performance, but jointly use of GNB and WFS provides higher and more stable outcomes (77.33\% and 88.57\% of accuracy) (Figure~\ref{fig:GNB}). Application of WFS to Australian data (Figure~\ref{fig:DT}), and GS to German data  is more appropriate in terms of DT's classification ability. Although there is a feature selection mechanism in DT (called embedded method), WFS has a favorable impact. Three different approaches contribute significantly to the effectiveness of RF algorithm, but it is seen that while only GS is applied, better results are achieved compared to the others 80.67\% and 86.67\% of accuracy (Figure~\ref{fig:RF}). While analyzing with XGB, WFS has a positive impact on Australian data and a negative effect on German data. On the other hand, it is found out that GS is the best method on both data sets and yields 81.67\% and 89.05\% of accuracy (Figure~\ref{fig:XGB}).

\paragraph{Influence of transformations}For SVM the most successful technique is the standard scale on Australian data (88.10\% of accuracy). Nevertheless, the case when none of the scale techniques is applied gives the highest accuracy rate in German data (79.33\% of accuracy), however requires the longest computation time. For this reason, natural log transformation is more appropriate for SVM on German set (78.67\% of accuracy) (Figure~\ref{fig:SVM}). For KNN algorithm, Min-Max is more effective for Australian data (89.05\% of accuracy), and Standard scale for German data (Figure~\ref{fig:KNN}). It is found that Standard scale is the best option on both dataset with accuracy rates (82.00\% and 87.62\% of accuracy) for MLP. Box-Cox can be taken into account for an option on Australian data. However, the worst results for MLP are obtained with natural log on Australian data, and Min-Max on the German data (Figure~\ref{fig:MLP}). As it can be seen from Figure~\ref{fig:LR}, LR employed to both sets by implementing Min-Max and Standard scale give the highest scores. The Standard scale is more suitable, but natural log is more incompatible technique for Australian data. For GNB, the results indicate that Min-Max leads to a more accurate score on Australian data (88.10\% of accuracy), while Standard scale is not as efficient as the other techniques. On German data, in GNB the highest outcomes are achieved with Box-Cox and Standard techniques (Figure~\ref{fig:GNB}). Although in literature~\citep{zheng2018feature,raschka2017python,trevor2009elements} data transformation techniques are told to be ineffective on RF, DT and XGB tree based algorithms, we show that our findings support this on DT and XGB algorithms (Figures~\ref{fig:DT} and~\ref{fig:XGB},  respectively). However, contrary to the literature, we show that scaling techniques have an impact on RF algorithm as can be seen in Figure ~\ref{fig:RF}.   

\begin{table}[h]
	\caption{Selected Features for German credit data set using ML methods}
	\counterwithout{table}{section}
	\centering
	\scalebox{0.8}{
		\begin{tabular}{|l|l|l|}
			\hline
			\textbf{Algorithms}    & \textbf{BAL}                        & \textbf{CSA}                        \\ \hline
			DT          & 1,4,8,9,12                    & 1,4,8,9,12                    \\ \hline
			KNN     & 3,8,9,10,12,13,14              & 4,8,9,10,12,14                 \\ \hline
			LR    & 4,5,6,8,10,14                  & 1,2,3,4,5,6,8,9,10,12          \\ \hline
			MLP  & 1,2,3,4,5,6,7,8,9,11,12         & 1,2,3,4,5,6,7,8,9,11,12,13,14   \\ \hline
			GNB            & 1,2,4,5,6,8,9,11,12,13        & 1,2,4,5,6,8,9,11,12,13        \\ \hline
			SVM & 1,2,3,4,6,7,8,9,10,13          & 4,6,8,9                       \\ \hline
			RF          & 1,2,3,4,5,6,8,9,10,11,12,13,14 & 1,2,3,4,5,6,8,9,10,11,12,13,14 \\ \hline
			XGB                & 1,4,5,7,8,9,10,11,12,13        & 1,4,5,7,8,9,10,11,12,13        \\ \hline
	\end{tabular}}
	\label{tab:Feat}
\end{table}

\paragraph{Impact of feature selection method} It is observed that some of the attributes are commonly selected by all ML algorithms. We classify them in two groups: the \textbf{Best Accuracy List} (BAL) and the \textbf{Commonly Selected Attributes} (CSA). The maximum accuracy achieving attributes are listed in the BAL sets, whereas, in three trials of feature selections, the attributes which appear more than once are categorized as CSA. These indicators are summarized in Table \ref{tab:Feat} only for German data set, as it makes sense to interpret them due to the availability on the definition of attributes whose code information can be found in Table \ref{tab:abc}. The features in German data which are contained at least five of the CSA sets are Checking Account, Duration in month, Credit History, Credit Amount, Savings Account, Employment Since, Present Resident, Foreign Worker, Purpose-New Car, Purpose-Used Car, Purpose-Domestic Appliances, Co-Applicant. On the other hand, Checking Account, Duration in month, Credit History, Savings Account, Employment Since, Foreign Worker, Purpose-New Car, Purpose-Used Car, Purpose-Domestic Appliances, Co-Applicant, Skilled Employee-Official are included in more than four of the BAL sets. As the definition of the attributes are not known, we avoid comment on the selection of attributes in Australian data set, however these can be obtain similarly.	

\paragraph{The computing time} We compare the run-time of each ML-feature-scaling combination for both data sets whose results are listed in Tables \ref{tab:execute1} and \ref{tab:execute3}. On German data, the case of no scaling in SVM with GS gives the highest accuracy rate, but takes the longest execution time of about 6 hours, compared to the average computational time of the other transformations which is 8 min. and 34 sec. On the other hand, in Australian data, the run time in DT GNB, KNN, RF and LR ends on average in seconds due to the few numbers of attribute and observations. It should also be marked that the programs are run in PC-i5 environment using Python 3.6.3.

\paragraph{Performance against unbalanced data} Most of the time there are relatively few observations from one class in real world datasets. With regard to determine more suitable algorithm on unbalanced dataset, we consider German data (30-70\%) since it is more unbalanced in terms of class distribution compared to Australian data set (45-55\%). To validate performance of the algorithms, we consider average balanced error rates (BER), equally weighted type I and type II errors, of all algorithm states. Figure \ref{fig:tyiii} indicates that XGB is more appropriate (29\%) for this purpose, and DT is the least one (38\%).

\begin{figure}[h!]
	\centering
	\caption{Illustration of average BER scores for German data}
	\includegraphics[width=0.5\linewidth]{Fig1-eps-converted-to}
	\label{fig:tyiii}
\end{figure}

\begin{table*}[h]
	\counterwithout{table}{section}
	\centering
	\caption{Comparison of the findings with respect to related literature (in \%)}
	\scalebox{0.75}{%
		\begin{tabular}{llcccccccc}                                                                                                                                                                                                                               \\ \hline
			& \textbf{}                                 & \multicolumn{1}{l}{\textbf{}}       & \multicolumn{3}{c}{\textbf{German Data}}                                                       & \multicolumn{1}{l}{}      & \multicolumn{3}{c}{\textbf{Australian Data}}                                 \\ \hline
			& \multicolumn{1}{c}{\textbf{Author}} & \textbf{ML Method}                     & \textbf{AR}        & \textbf{Type I }         & \textbf{Type II}              & \textbf{ML}           & \textbf{AR}           & \textbf{Type I} & \textbf{Type II} \\
			& \textbf{Our Study}  & MLP\&G.S  & 82.00   & 40.00 & 8.57 & XGB \& G.S.   & 89.05  & 11.96 & 9.67  \\ \hline
			& Shen et al. (2019) & SMOTE & 78.70 & 30.00 & 17.57 & SMOTE & 90.58 & 4.8 & 13.05 \\		
			& Biswas et al. (2018) & RxNCM & 66.00 & 40.00 & NA & RxNCM & 78.26 & 11.11 & NA \\		
			& Ramasamy et al. (2018) & OGD-RBM  & 76.5  & 40.00  & 17.00  & OGD-RBM  & 88.49  & 8.00  & 14.00  \\		
			& Xia et al. (2017) & Bagging NN  & 76.01  & 49.67  & 12.98  & XGB  & 87.81  & 13.92  & 10.80  \\		
			& Liang et al. (2015) & SVM  & 76.30   & 49.80   & NA  & NB+PSO   & 85.86  & 12.41  & NA  \\		
			& Marques et al. (2012) & MLP+Adaboost  & 71.50  & 45.00 & NA & Adaboost+NB  & 79.57 & 3.00  & NA   \\		
			& Yu et al. (2010)  & LR   & 75.47   & 15.30    & NA  & RBF SVM  & 85.48  & 3.30   & NA \\
			& Tsai (2009)  & MLP+FA   & 78.76   & 48.69    & 10.66  & MLP+PCA  & 89.93  & 11.53   & 7.9  \\ \hline
			\multicolumn{10}{l}{FA:Factor Analysis, \  SMOTE:The synthetic minority over-sampling technique, \ PSO:Partial Swarm Optimization}\\
			\multicolumn{10}{l}{RxNCM:Rule Extraction from Neural Network using Classified  and  Misclassified  data, \ PCA:Principal Component Analysis} \\
			\multicolumn{10}{l}{OGD-RBM:Online Generative Discriminative Restricted Boltzmann Machine \ \ NA:Not Available}
	\end{tabular}}
	\label{tab:lit}
\end{table*}

To emphasize more on the contribution in this study, the comparisons to the literature at which ML methods are employed on the same data sets and AR, Type I and II error ratios are shared, are summarized in Table \ref{tab:lit}. More than twenty recently published studies, the ones which present the highest accuracy and the smallest error rates are chosen for comparison. It is shown that with our analyses, MLP with GS on German data set and XGB with GS on Australian credit data come up with preferable accuracy indicators (AR, Type I and II errors) than the selected studies by a slight difference in MLP and PCA case for Australian data set proposed by ~\citep{tsai2009feature}. For the same data set, a Type I error of 4.8\% is found in SMOTE appears to be very low, it is seem to cause Type II error rate higher~\citep{shen2019novel}. For German data set our analysis gives higher accuracy, and lower Type II error. Although, work of ~\citep{yu2010comparative} gives quite low Type I error of 15.30\%, it also has one of the lowest AR (75.47\%)  and study of~\citep{shen2019novel} gives good results in terms of a type I error, but it remains high in terms of type two compared to other and our studies.

\begin{table}[h!]
	\centering
	\caption{Computation times in German dataset}
	\counterwithout{table}{section}
	\label{tab:execute1}
	\scalebox{0.6}{%
		\begin{tabular}{lcccccccc}
			\hline
			\multicolumn{1}{c}{\textbf{Scale}} & \textbf{XGB} & \textbf{RF} & \textbf{DT} & \textbf{GNB} & \textbf{SVM} & \textbf{MLP} & \textbf{KNN} & \textbf{LR} \\ \hline
			\multicolumn{9}{c}{Grid Search} \\ \hline
			Without & 0:00:30 & 0:01:12 & 0:00:01 & 0:00:08 & 5:40:49 & 0:01:07 & 0:00:09 & 0:00:35 \\
			Standard & 0:00:30 & 0:01:14 & 0:00:01 & 0:00:01 & 0:09:38 & 0:01:11 & 0:00:02 & 0:00:23 \\
			Min-Max & 0:00:30 & 0:01:23 & 0:00:01 & 0:00:01 & 0:07:09 & 0:01:11 & 0:00:02 & 0:00:23 \\
			Box-Cox & 0:00:30 & 0:01:13 & 0:00:01 & 0:00:01 & 0:08:24 & 0:01:10 & 0:00:02 & 0:00:38 \\
			natural log & 0:00:30 & 0:01:12 & 0:00:01 & 0:00:01 & 0:06:45 & 0:01:11 & 0:00:02 & 0:00:27 \\
			Min-Max (All At.) & 0:00:30 & 0:01:13 & 0:00:01 & 0:00:01 & 0:08:04 & 0:01:17 & 0:00:02 & 0:00:11 \\
			Standard (All At.) & 0:00:30 & 0:01:13 & 0:00:01 & 0:00:01 & 0:11:23 & 0:01:03 & 0:00:02 & 0:00:07 \\ \hline
			\multicolumn{9}{c}{Wrapper Feature Selection} \\ \hline
			Without & 0:10:07 & 0:02:36 & 0:00:36 & 0:00:28 & 0:08:37 & 0:25:33 & 0:00:42 & 0:03:04 \\
			Standard & 0:10:08 & 0:02:39 & 0:00:36 & 0:00:27 & 0:06:15 & 0:41:48 & 0:00:43 & 0:02:18 \\
			Min-Max & 0:10:06 & 0:02:29 & 0:00:36 & 0:00:27 & 0:03:50 & 0:37:21 & 0:00:43 & 0:02:45 \\
			Box-Cox & 0:09:58 & 0:02:42 & 0:00:36 & 0:00:29 & 0:06:42 & 0:34:06 & 0:00:43 & 0:03:02 \\
			natural log & 0:09:58 & 0:02:30 & 0:00:36 & 0:00:27 & 0:04:23 & 0:30:28 & 0:00:43 & 0:02:33 \\
			Min-Max (All At.) & 0:10:07 & 0:02:56 & 0:00:36 & 0:00:27 & 0:01:31 & 0:40:19 & 0:00:42 & 0:01:23 \\
			Standard (All At.) & 0:09:58 & 0:02:33 & 0:00:36 & 0:00:27 & 0:01:41 & 0:42:16 & 0:00:44 & 0:00:55 \\ \hline
		\end{tabular}%
	}
\end{table}

\begin{table}[h!]
	\centering
	\caption{Computation times in Australian dataset}
	\counterwithout{table}{section}
	\label{tab:execute3}
	\scalebox{0.6}{%
		\begin{tabular}{lcccccccc}
			\hline
			\multicolumn{1}{c}{\textbf{Scale}} & \textbf{XGB} & \textbf{RF} & \textbf{DT} & \textbf{GNB} & \textbf{SVM} & \textbf{MLP} & \textbf{KNN} & \textbf{LR} \\ \hline
			\multicolumn{9}{c}{Grid Search} \\ \hline
			Without & 0:01:07 & 0:00:38 & 0:00:01 & 0:00:01 & 0:01:26 & 0:00:11 & 0:00:43 & 0:00:06 \\
			Standard & 0:01:07 & 0:00:44 & 0:00:01 & 0:00:01 & 0:01:26 & 0:00:22 & 0:01:03 & 0:00:08 \\
			Min-Max & 0:01:07 & 0:00:36 & 0:00:01 & 0:00:01 & 0:01:25 & 0:00:20 & 0:00:56 & 0:00:01 \\
			Box-Cox & 0:01:07 & 0:00:36 & 0:00:01 & 0:00:01 & 0:01:22 & 0:00:19 & 0:00:49 & 0:00:03 \\
			natural log & 0:01:07 & 0:00:38 & 0:00:01 & 0:00:01 & 0:00:13 & 0:00:19 & 0:00:40 & 0:00:03 \\
			Min-Max (All At.) & 0:01:07 & 0:00:37 & 0:00:01 & 0:00:01 & 0:01:37 & 0:00:21 & 0:01:04 & 0:00:01 \\
			Standard (All At.) & 0:01:07 & 0:00:36 & 0:00:01 & 0:00:01 & 0:01:20 & 0:00:22 & 0:01:07 & 0:00:01 \\ \hline
			\multicolumn{9}{c}{Wrapper Feature Selection} \\ \hline
			Without & 0:04:07 & 0:01:06 & 0:00:15 & 0:00:16 & 0:01:47 & 0:11:39 & 0:00:25 & 0:00:34 \\
			Standard & 0:04:07 & 0:01:09 & 0:00:15 & 0:00:13 & 0:01:37 & 0:10:53 & 0:00:30 & 0:00:22 \\
			Min-Max & 0:04:07 & 0:01:10 & 0:00:15 & 0:00:15 & 0:01:49 & 0:16:59 & 0:00:29 & 0:00:29 \\
			Box-Cox & 0:04:07 & 0:01:09 & 0:00:15 & 0:00:17 & 0:00:45 & 0:15:31 & 0:00:29 & 0:00:33 \\
			natural log & 0:04:07 & 0:01:01 & 0:00:15 & 0:00:18 & 0:01:38 & 0:12:08 & 0:00:25 & 0:00:35 \\
			Min-Max (All At.) & 0:04:07 & 0:01:12 & 0:00:15 & 0:00:15 & 0:00:58 & 0:13:37 & 0:00:29 & 0:00:21 \\
			Standard (All At.) & 0:04:07 & 0:01:07 & 0:00:15 & 0:00:16 & 0:01:41 & 0:13:35 & 0:00:27 & 0:00:19 \\ \hline
		\end{tabular}%
	}
\end{table}

\section{Concluding Comments}

This paper investigates the application of the eight ML methods on the two publicly available credit data sets; German \& Australian. The survey seeks which of these eight ML methods is superior to others to capture the default risk of consumer under WFS methods with use of SFS and data transformation techniques. The accuracy of these algorithms are compared with respect to some important indicators, such as AUC, accuracy rate, Type I and Type II error rates. The main contributions of this research, we conclude that the accuracy and the performance of ML algorithms is varying with respect to the feature selection method, transformation (scaling), data size, types and its partition in to train sets. In general, GS is the most suitable method for all algorithms, while WFS is more appropriate for KNN and GNB because of the increase in their accuracy ratio compared to their default case (no scale, no feature selection or GS). When all the outcomes are considered, WFS gives consistent results as it leads to the selection of similar features. In terms of data transformation techniques, Standard and Min-Max scale methods lead better outcomes in general. We also illustrate that compared to the literature, which of those methods with proposed combinations should be chosen by practitioners.

\bibliographystyle{unsrtnat}
\bibliography{references}  %%% Uncomment this line and comment out the ``thebibliography'' section below to use the external .bib file (using bibtex) .

%%% Uncomment this section and comment out the \bibliography{references} line above to use inline references.
% \begin{thebibliography}{1}

% 	\bibitem{kour2014real}
% 	George Kour and Raid Saabne.
% 	\newblock Real-time segmentation of on-line handwritten arabic script.
% 	\newblock In {\em Frontiers in Handwriting Recognition (ICFHR), 2014 14th
% 			International Conference on}, pages 417--422. IEEE, 2014.

% 	\bibitem{kour2014fast}
% 	George Kour and Raid Saabne.
% 	\newblock Fast classification of handwritten on-line arabic characters.
% 	\newblock In {\em Soft Computing and Pattern Recognition (SoCPaR), 2014 6th
% 			International Conference of}, pages 312--318. IEEE, 2014.

% 	\bibitem{hadash2018estimate}
% 	Guy Hadash, Einat Kermany, Boaz Carmeli, Ofer Lavi, George Kour, and Alon
% 	Jacovi.
% 	\newblock Estimate and replace: A novel approach to integrating deep neural
% 	networks with existing applications.
% 	\newblock {\em arXiv preprint arXiv:1804.09028}, 2018.

% \end{thebibliography}

{\centering
	\begin{figure*}[b]
		\caption{Illustration of efficiency indicators for German and Australian data sets}
		\begin{tabular}{cc} \textbf{German Dataset} \ \ \ \ \ \ \ \ \ \ \ \ \ \ \ \ \ \ \ \ \ \ \ \ \ \ \ \  \textbf{Australian Dataset} & \\
			\begin{subfigure}[b]{\textwidth}
				\centering
				\caption{SVM}
				\includegraphics[width=0.450\linewidth]{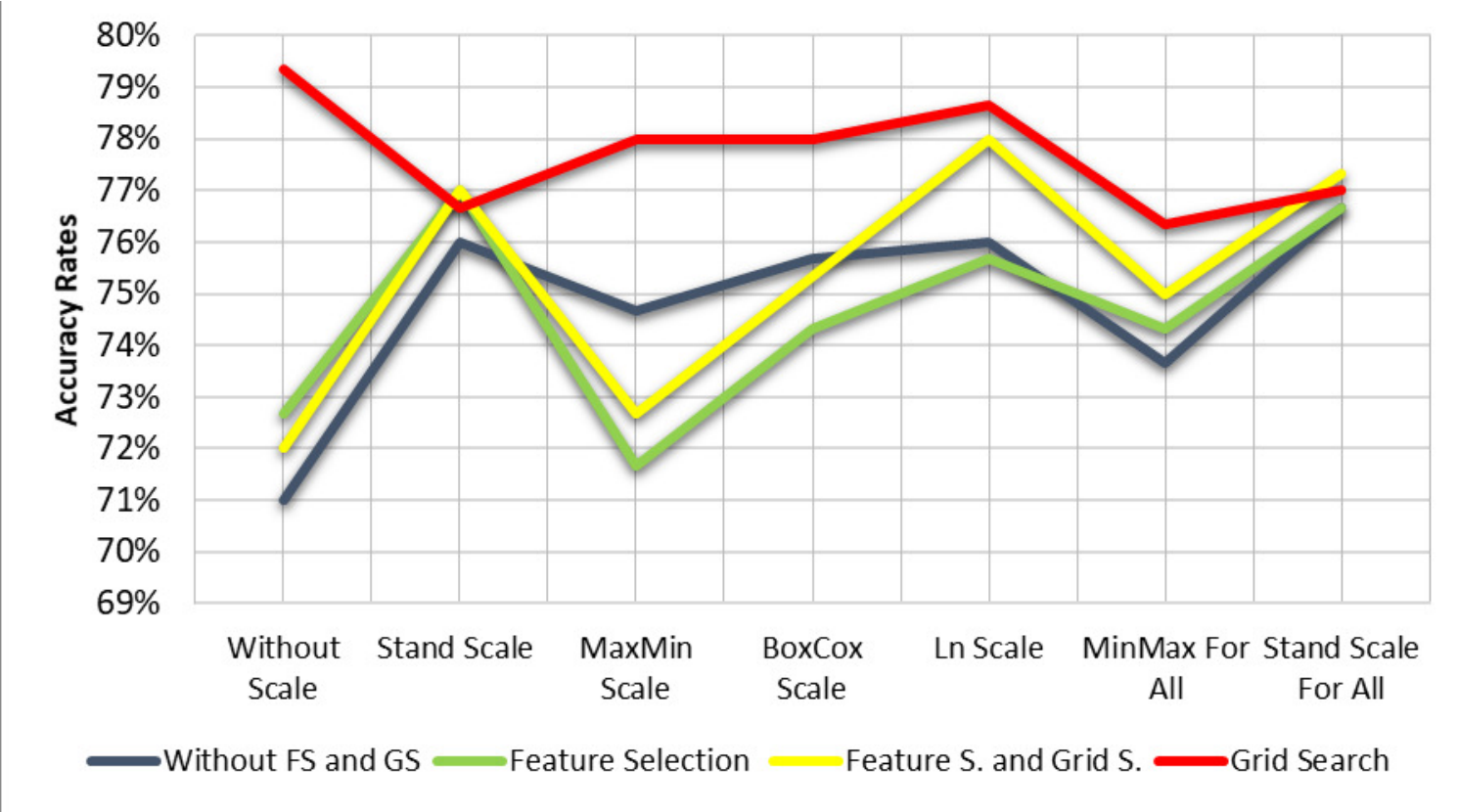}%
				\hfill
				\includegraphics[width=0.450\linewidth]{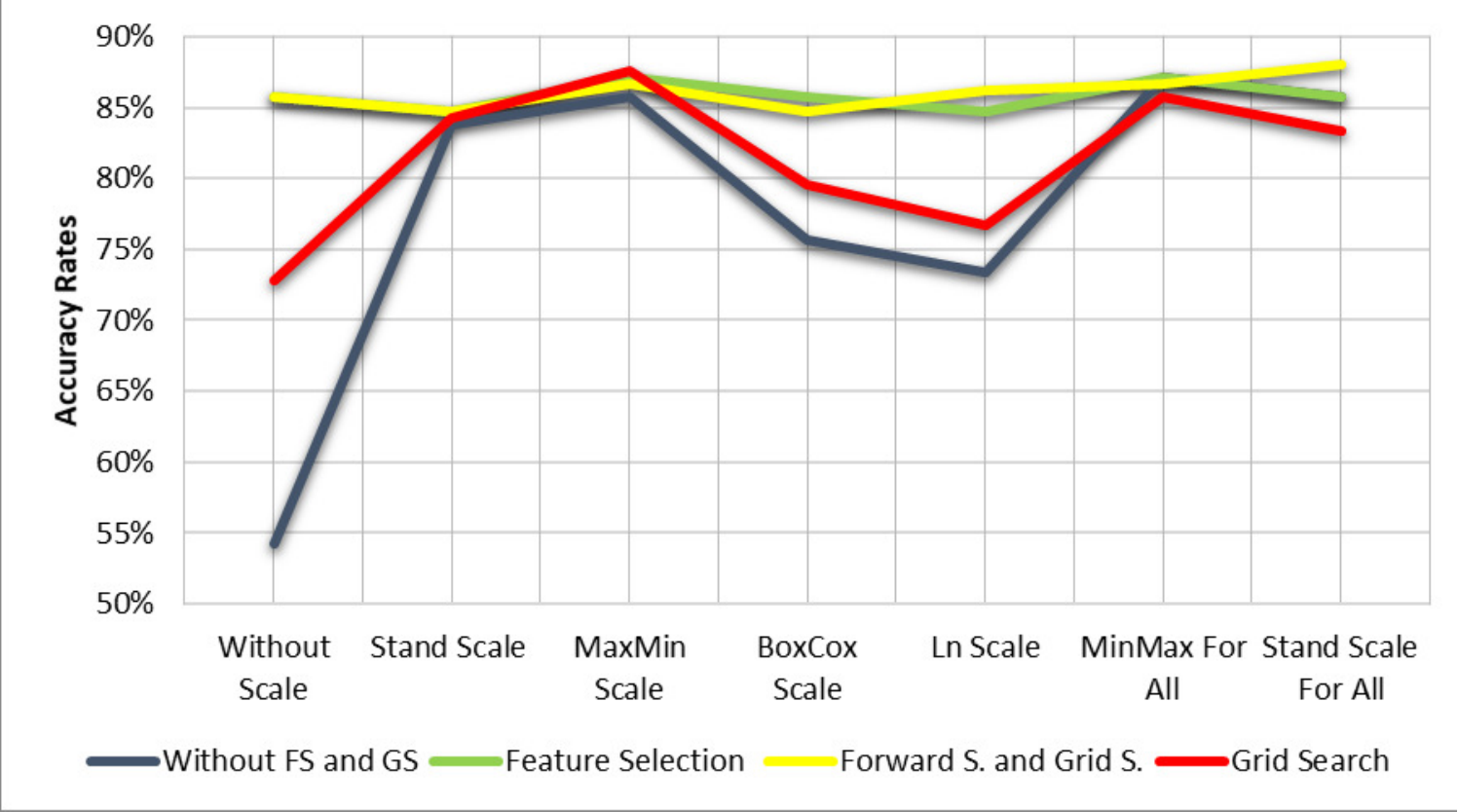}
				\label{fig:SVM}
			\end{subfigure} & \\
			\newline
			\begin{subfigure}[b]{\textwidth}
				\centering
				\caption{KNN}
				\includegraphics[width=0.450\linewidth]{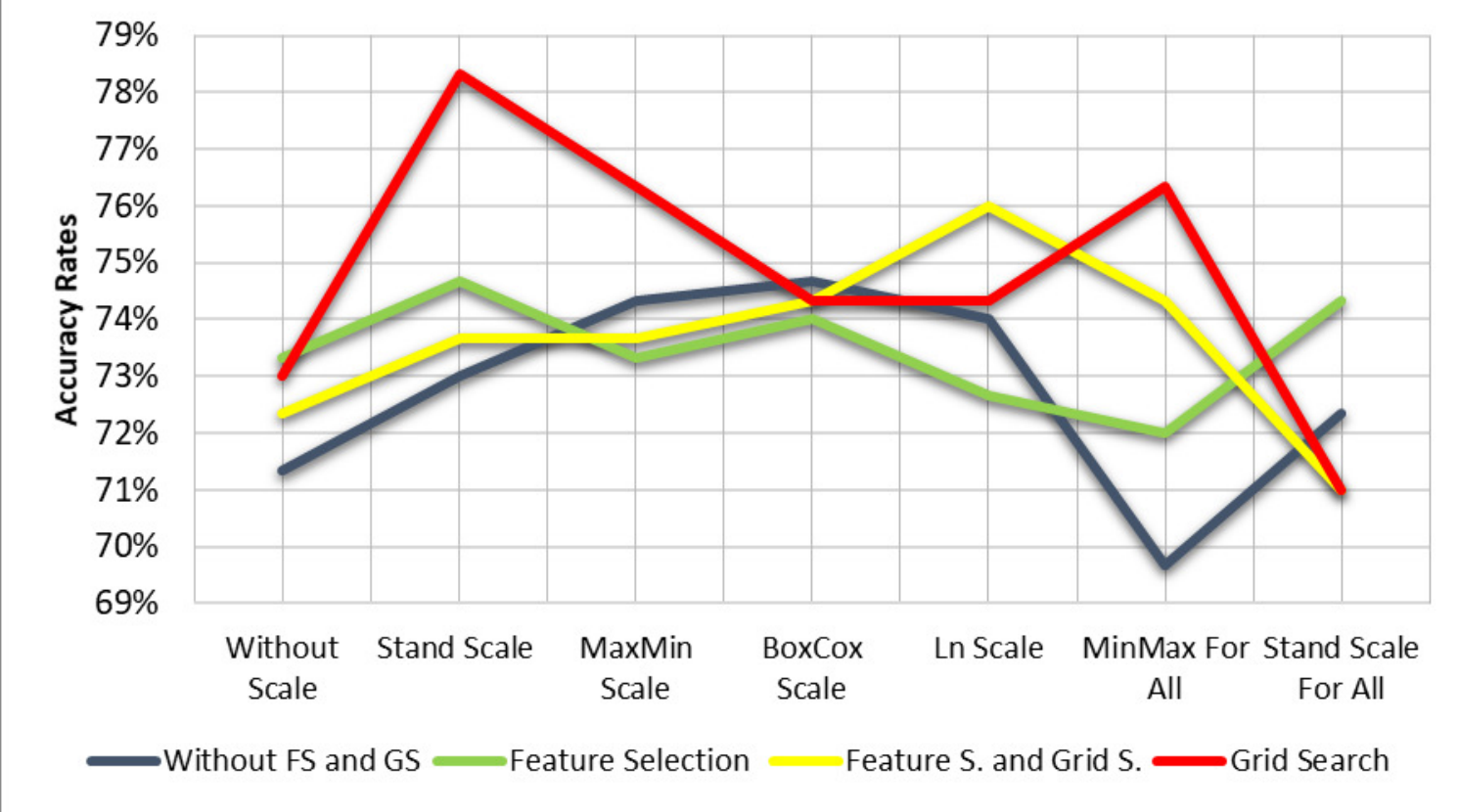}%
				\hfill
				\includegraphics[width=0.450\linewidth]{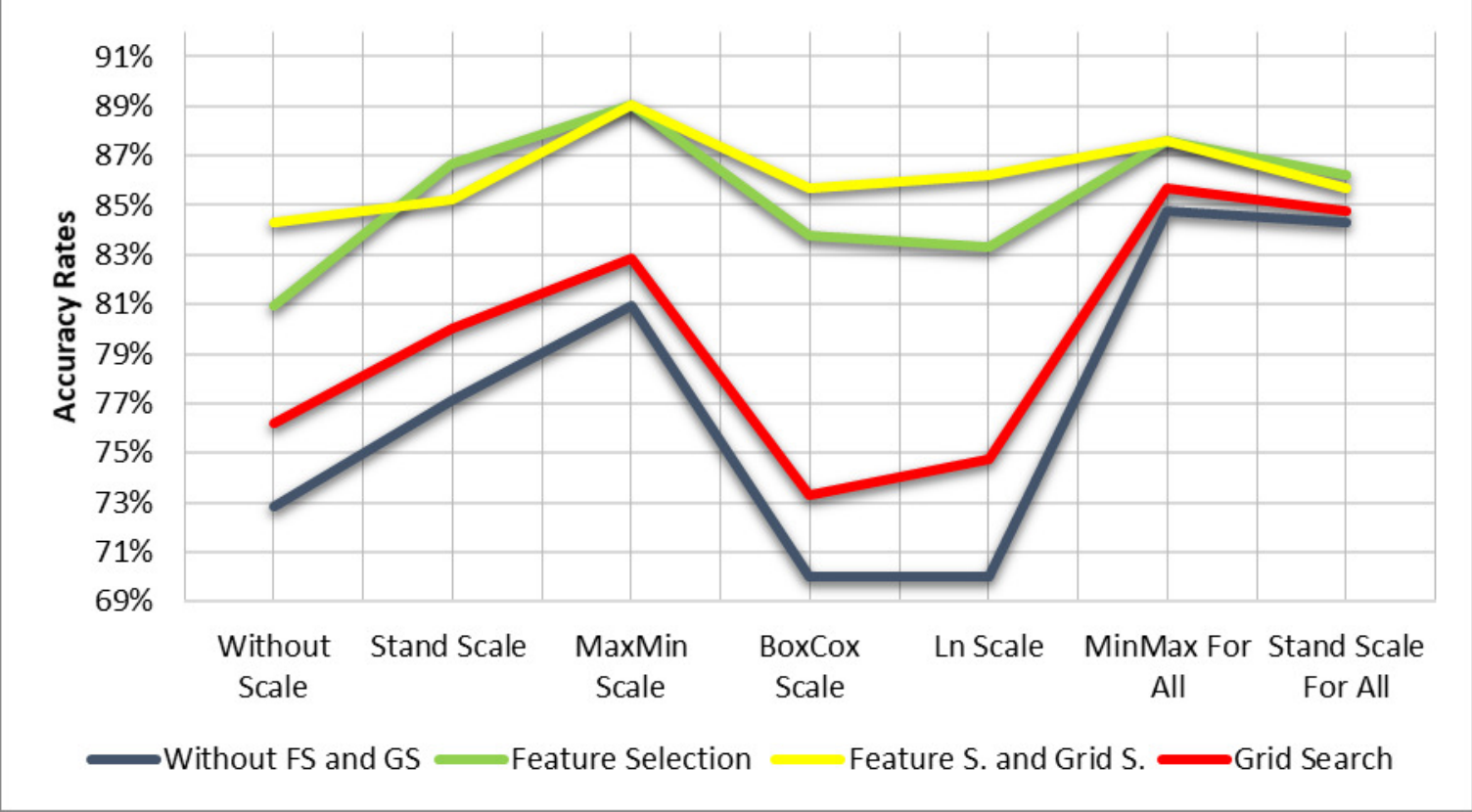}
				\label{fig:KNN}
			\end{subfigure} & \\
			\newline
			\begin{subfigure}[b]{\textwidth}
				\centering
				\caption{MLP}
				\includegraphics[width=0.450\linewidth]{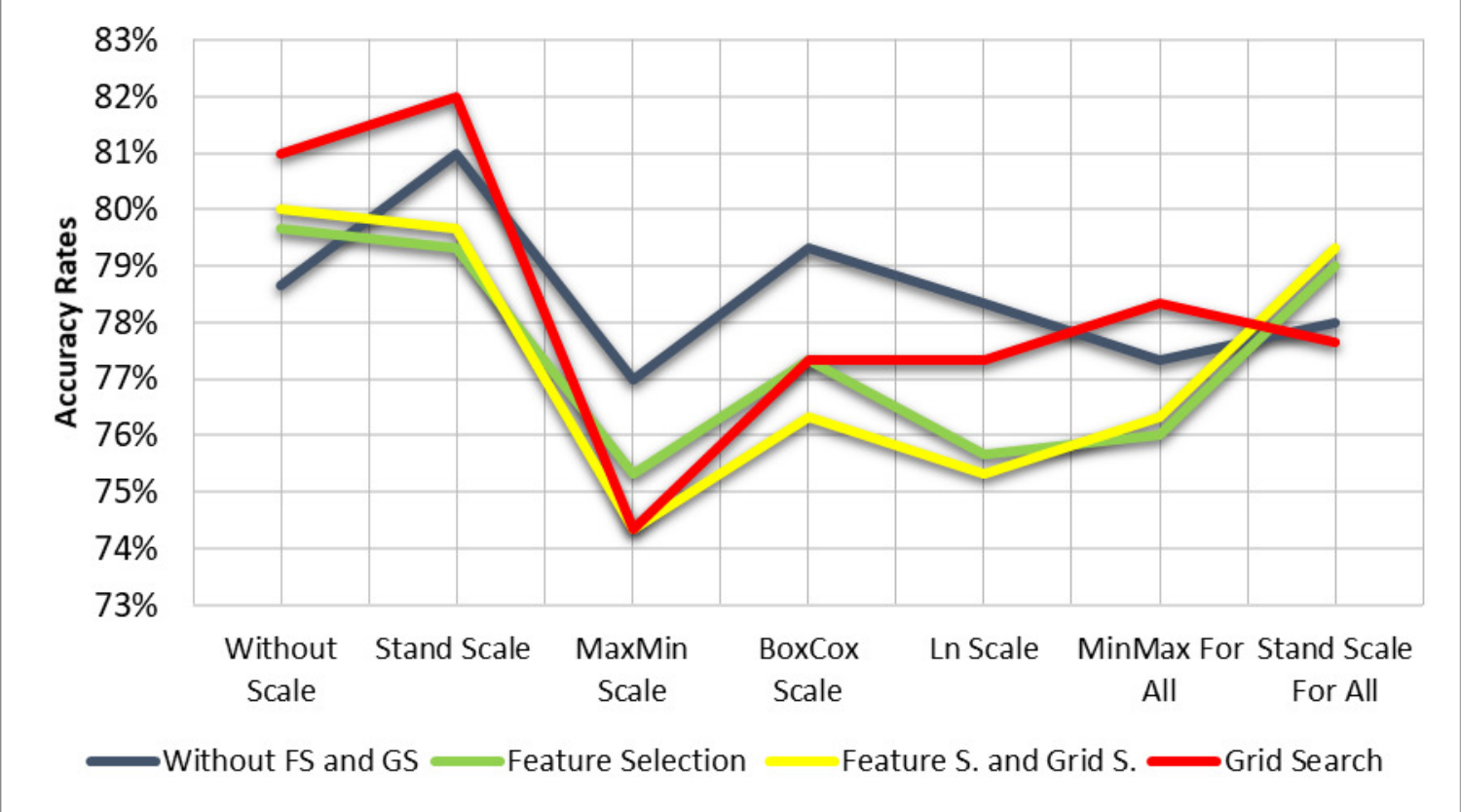}%
				\hfill
				\includegraphics[width=0.450\linewidth]{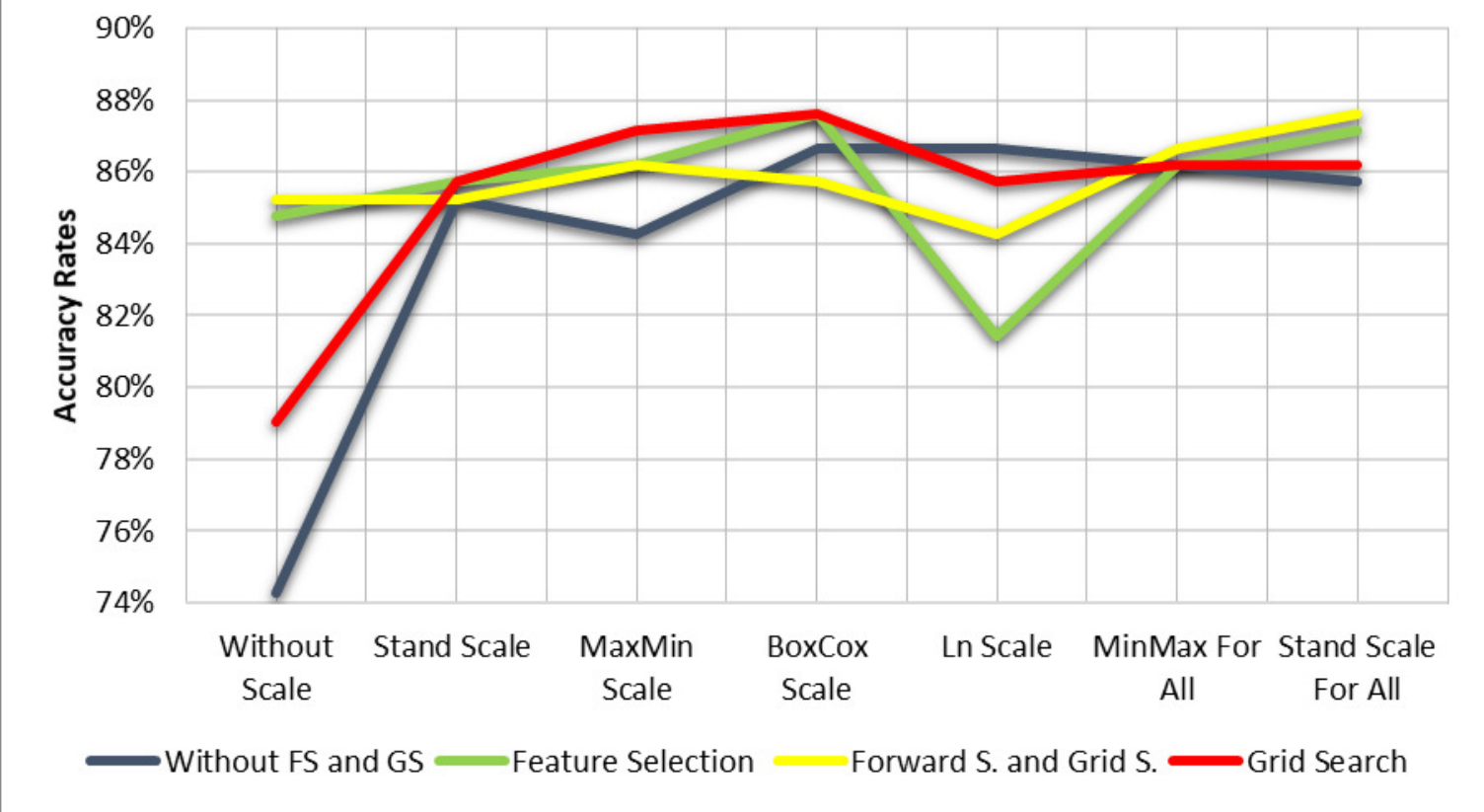}
				\label{fig:MLP}
			\end{subfigure} & \\
			\newline
			\begin{subfigure}[b]{\textwidth}
				\centering
				\caption{LR}
				\includegraphics[width=0.450\linewidth]{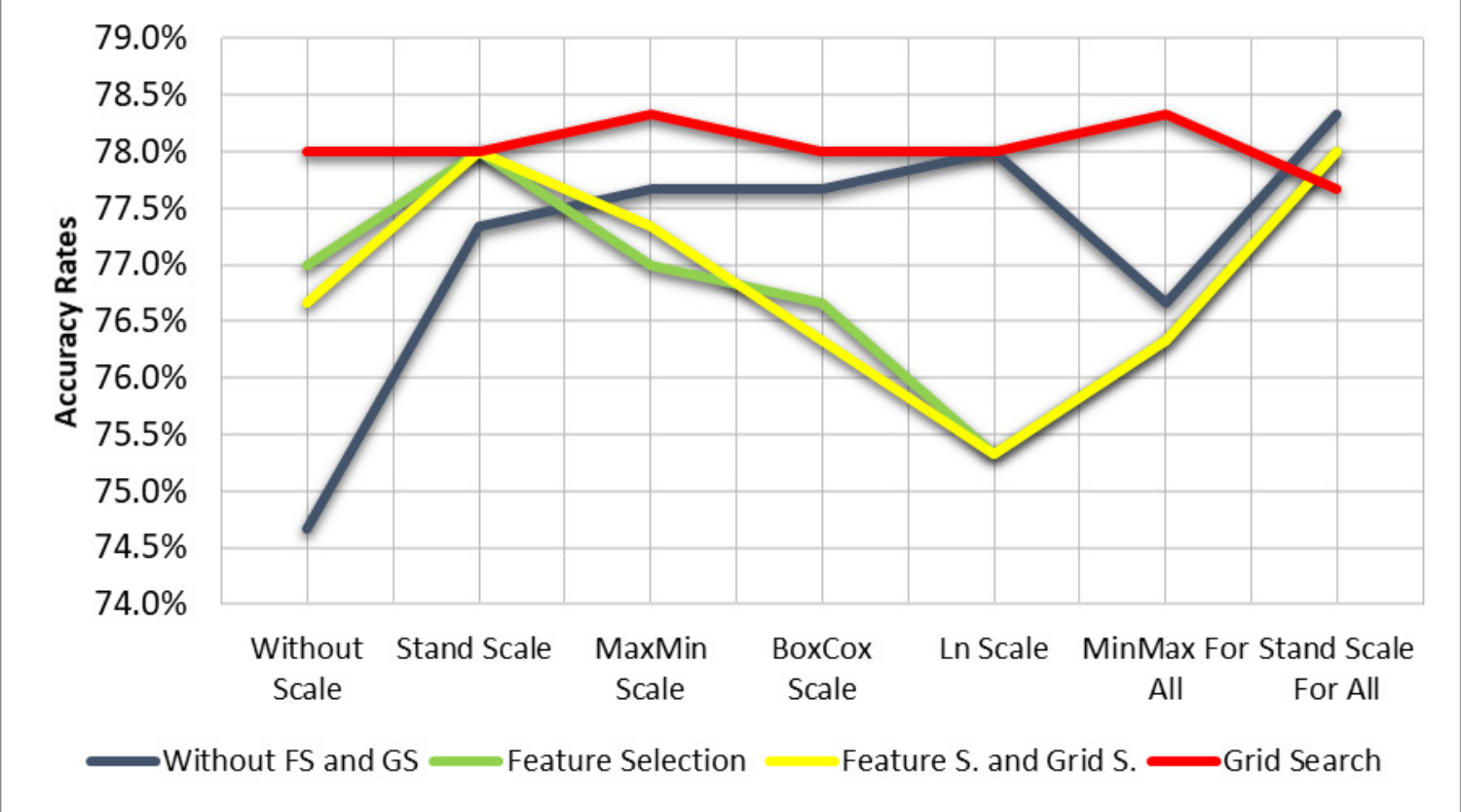}%
				\hfill
				\includegraphics[width=0.450\linewidth]{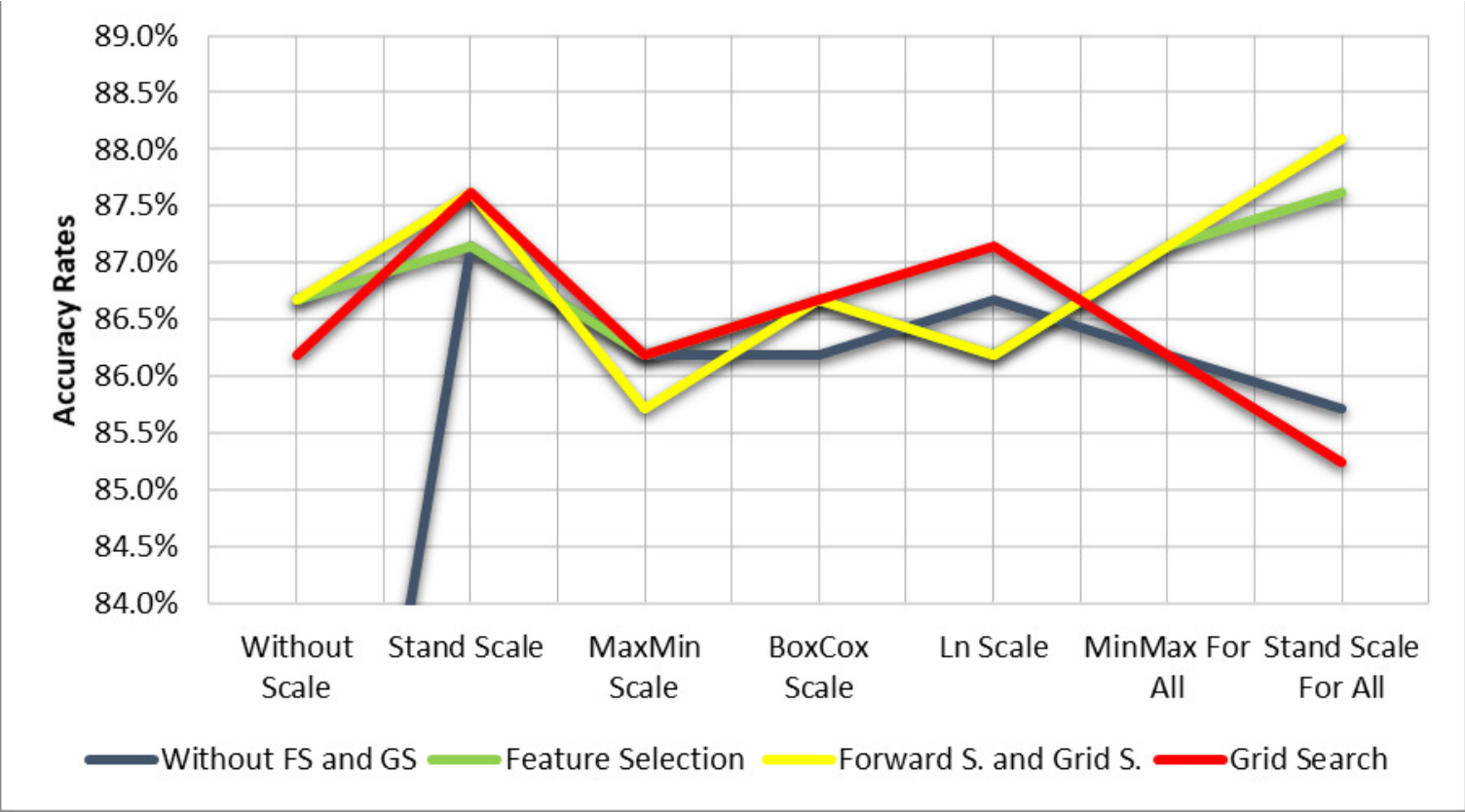}
				\label{fig:LR}
			\end{subfigure} & \\
		\end{tabular}
		\label{fig:mytable}
	\end{figure*}
}

{\centering
	\begin{figure*}[h]\ContinuedFloat
		\caption{Illustration of efficiency indicators for German and Australian data sets}
		\begin{tabular}{cc} \textbf{German Dataset} \ \ \ \ \ \ \ \ \ \ \ \ \ \ \ \ \ \ \ \ \ \ \ \ \ \ \ \  \textbf{Australian Dataset} & \\
			\begin{subfigure}[b]{\textwidth}
				\centering
				\caption{GNB}
				\includegraphics[width=0.450\linewidth]{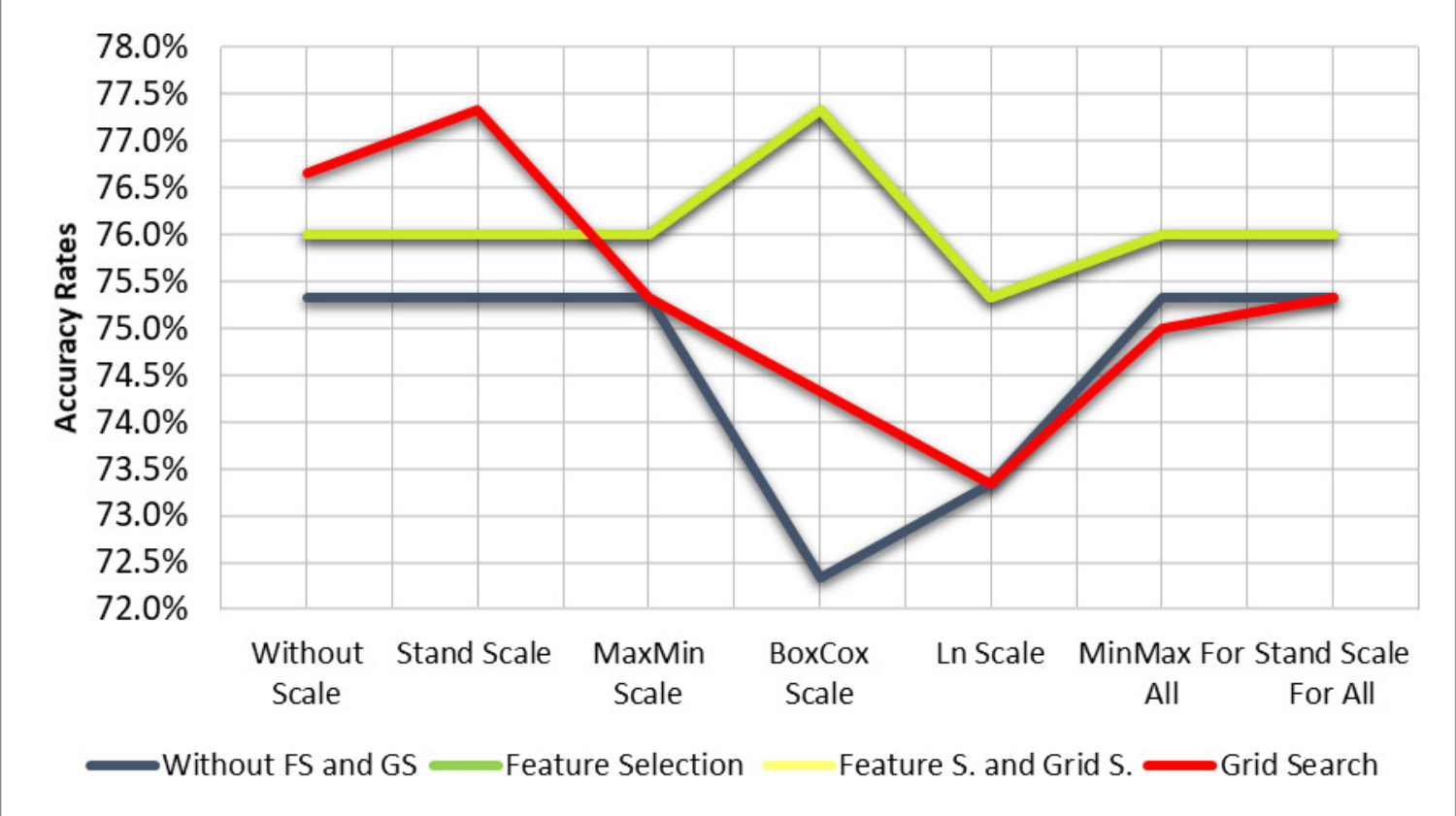}%
				\hfill
				\includegraphics[width=0.450\linewidth]{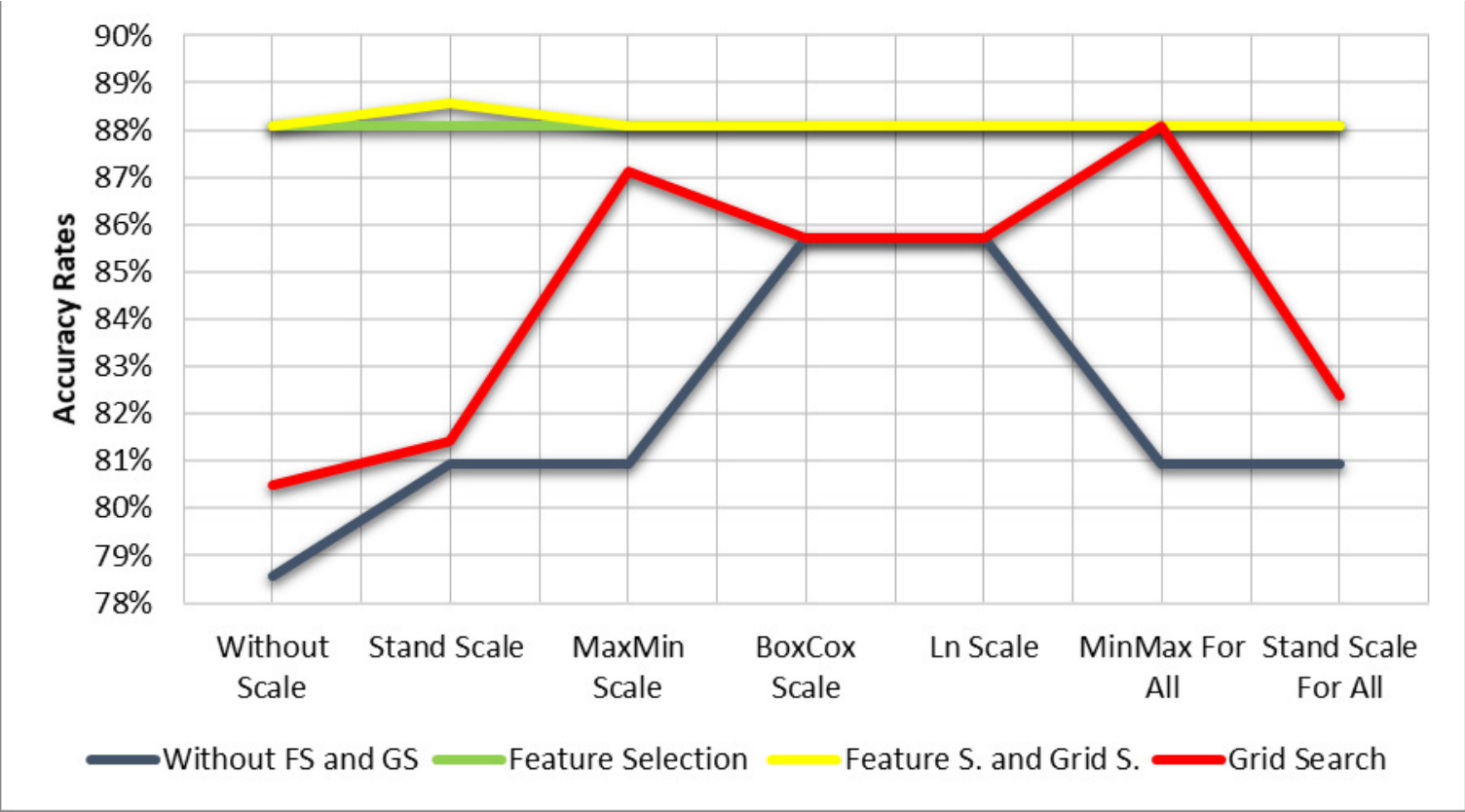}
				\label{fig:GNB}
			\end{subfigure} & \\
			\newline
			\begin{subfigure}[b]{\textwidth}
				\centering
				\caption{DT}
				\includegraphics[width=0.450\linewidth]{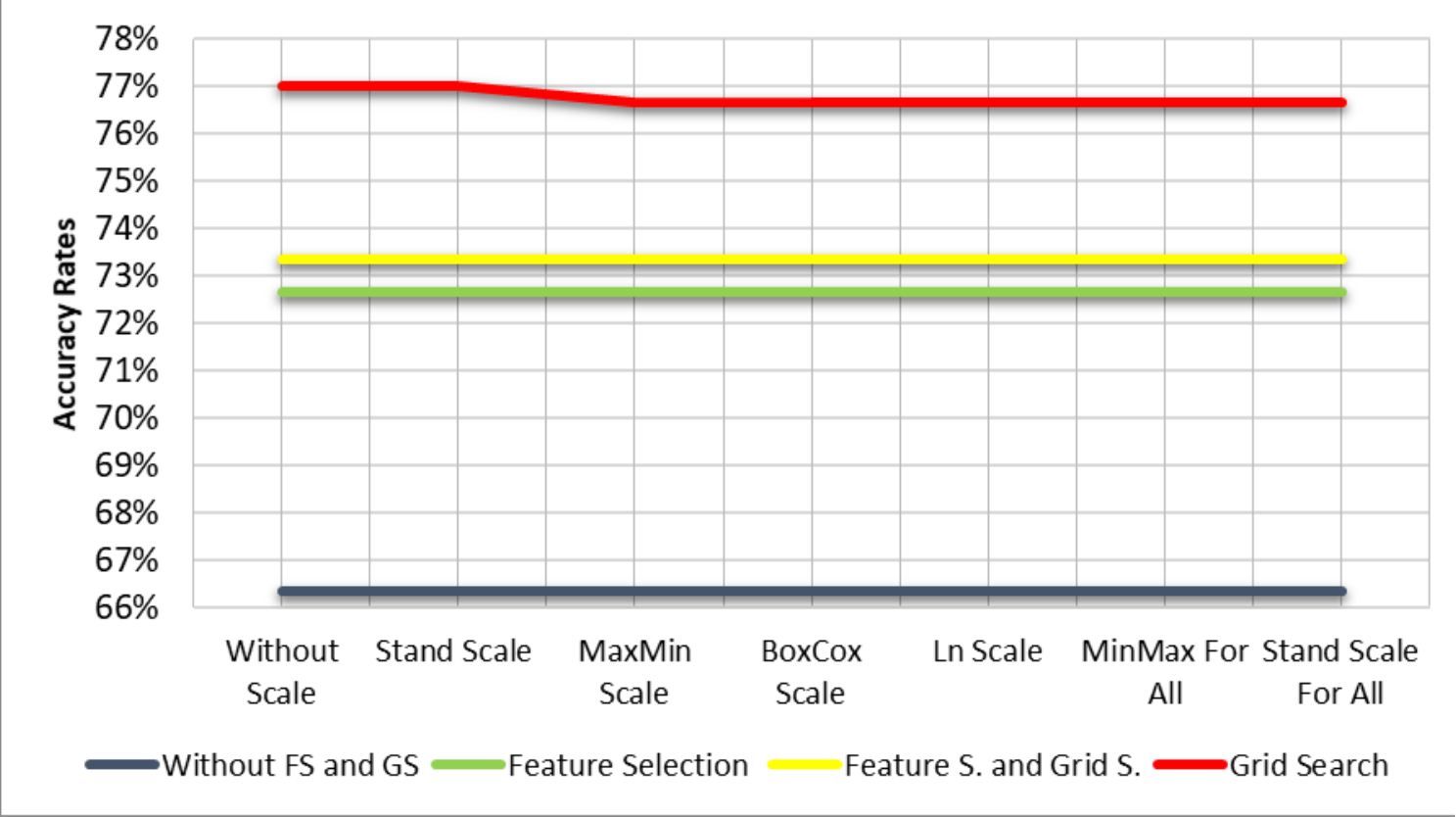}%
				\hfill
				\includegraphics[width=0.450\linewidth]{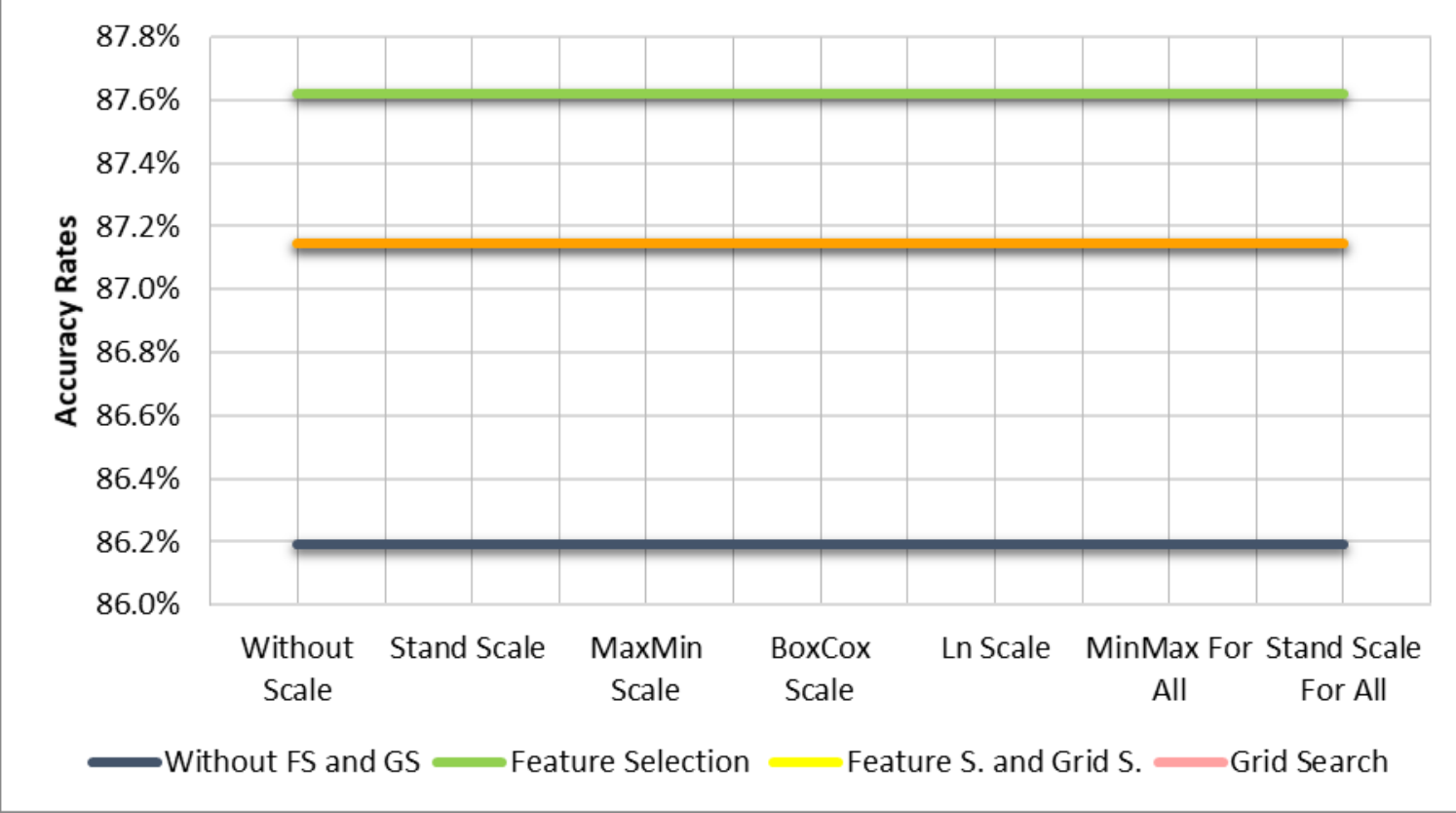}
				\label{fig:DT}
			\end{subfigure} & \\
			\newline
			\begin{subfigure}[b]{\textwidth}
				\centering
				\caption{RF}
				\includegraphics[width=0.450\linewidth]{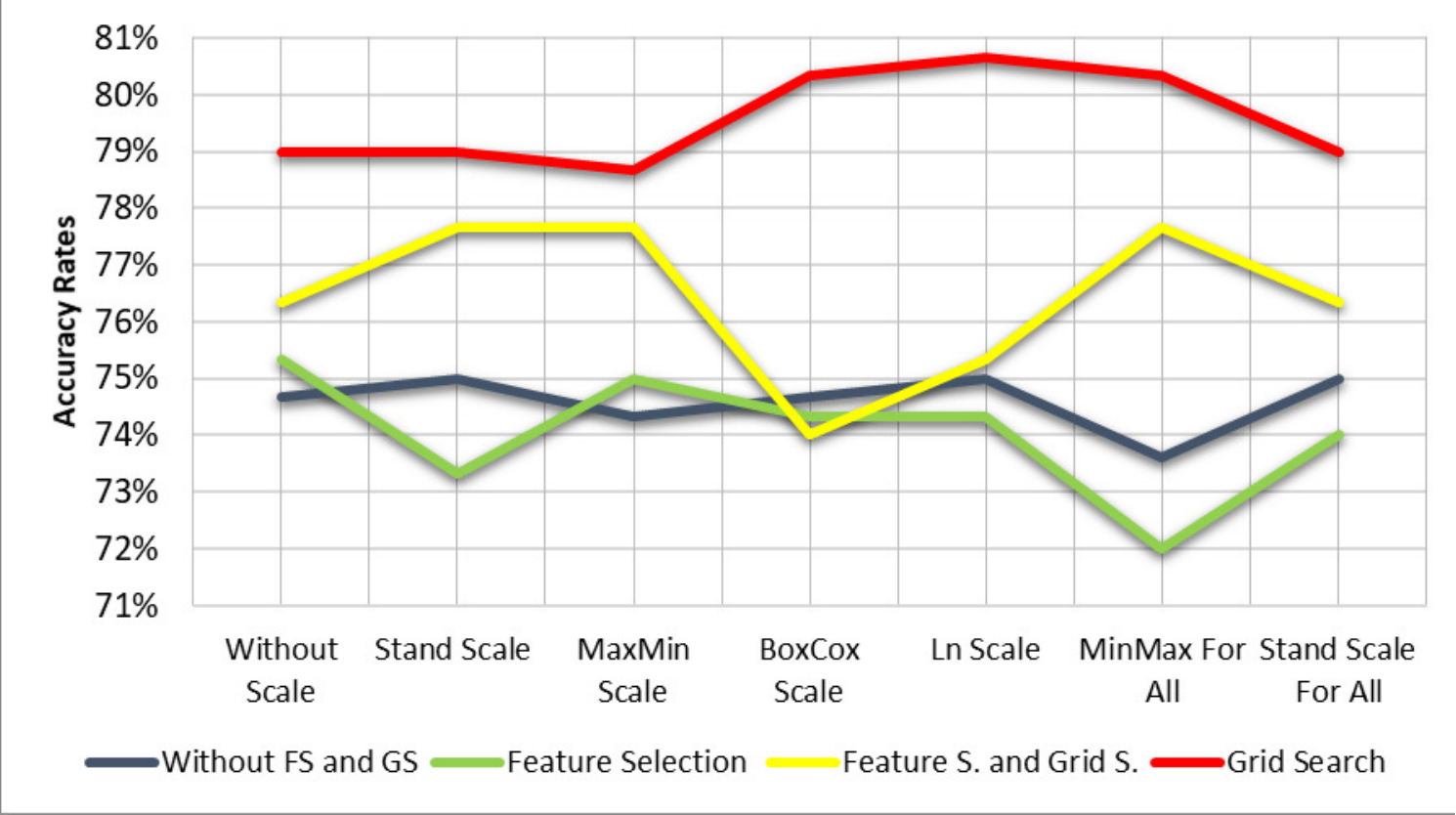}%
				\hfill
				\includegraphics[width=0.450\linewidth]{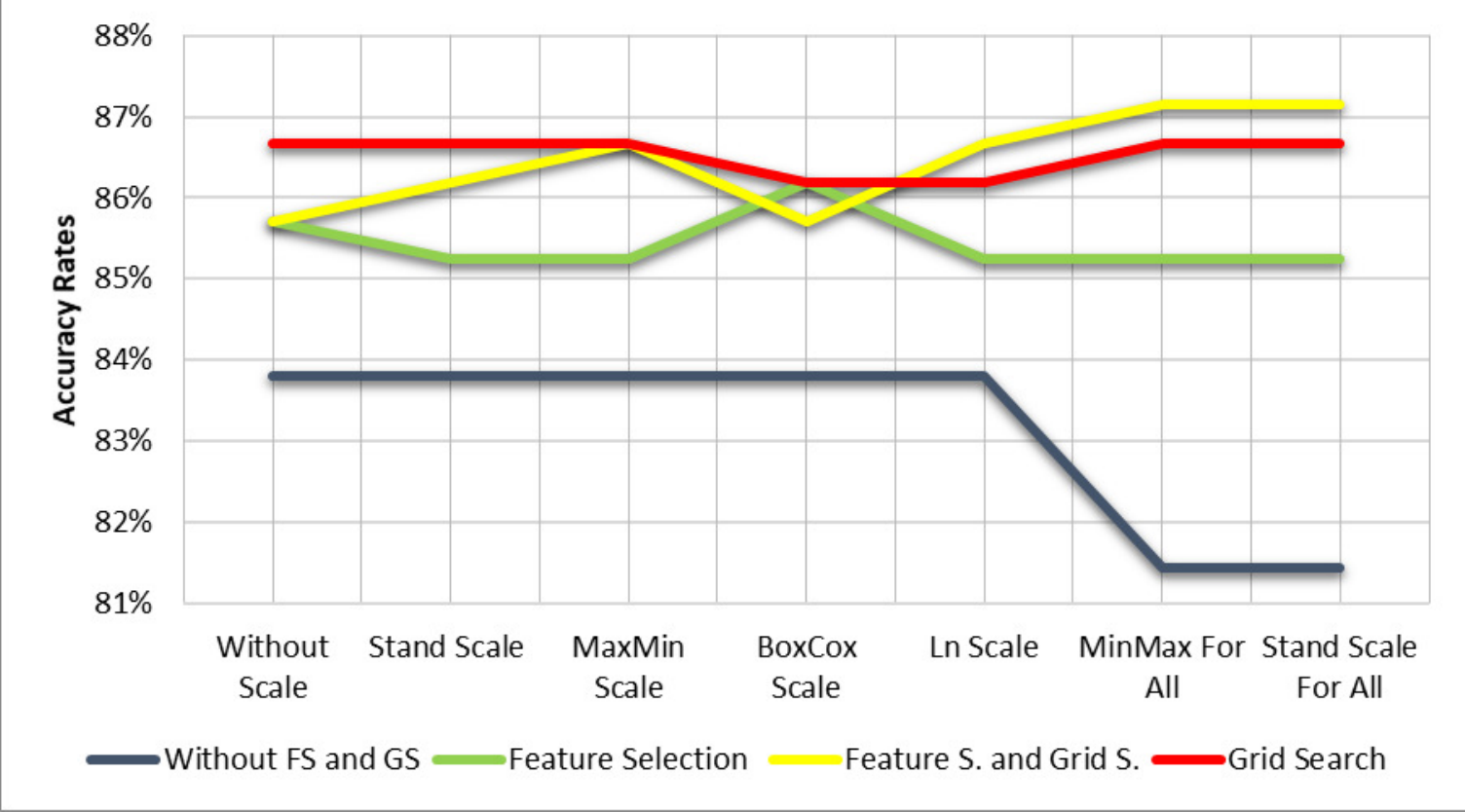}
				\label{fig:RF}
			\end{subfigure} & \\
			\newline
			\begin{subfigure}[b]{\textwidth}
				\centering
				\caption{XGB}
				\includegraphics[width=0.450\linewidth]{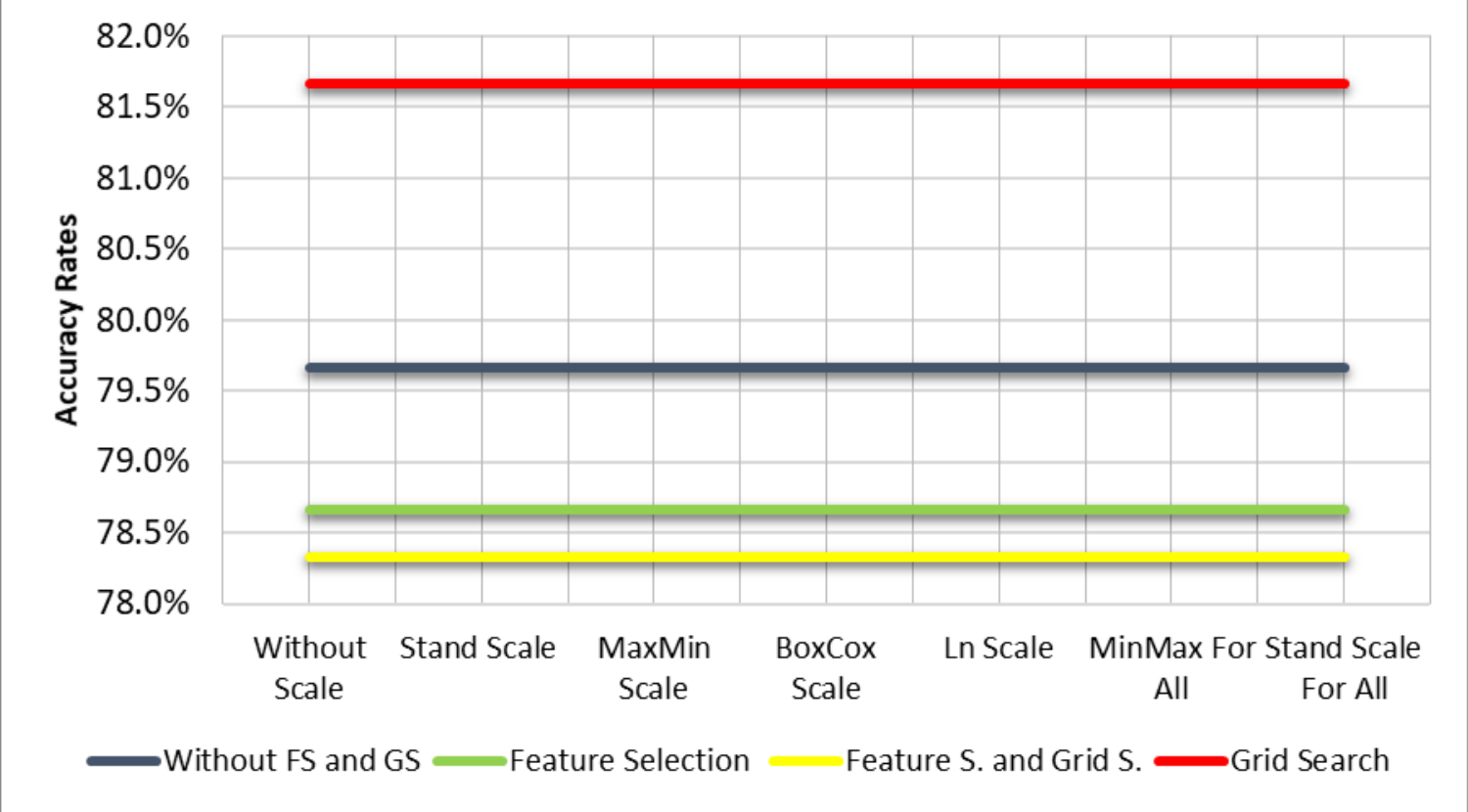}%
				\hfill
				\includegraphics[width=0.450\linewidth]{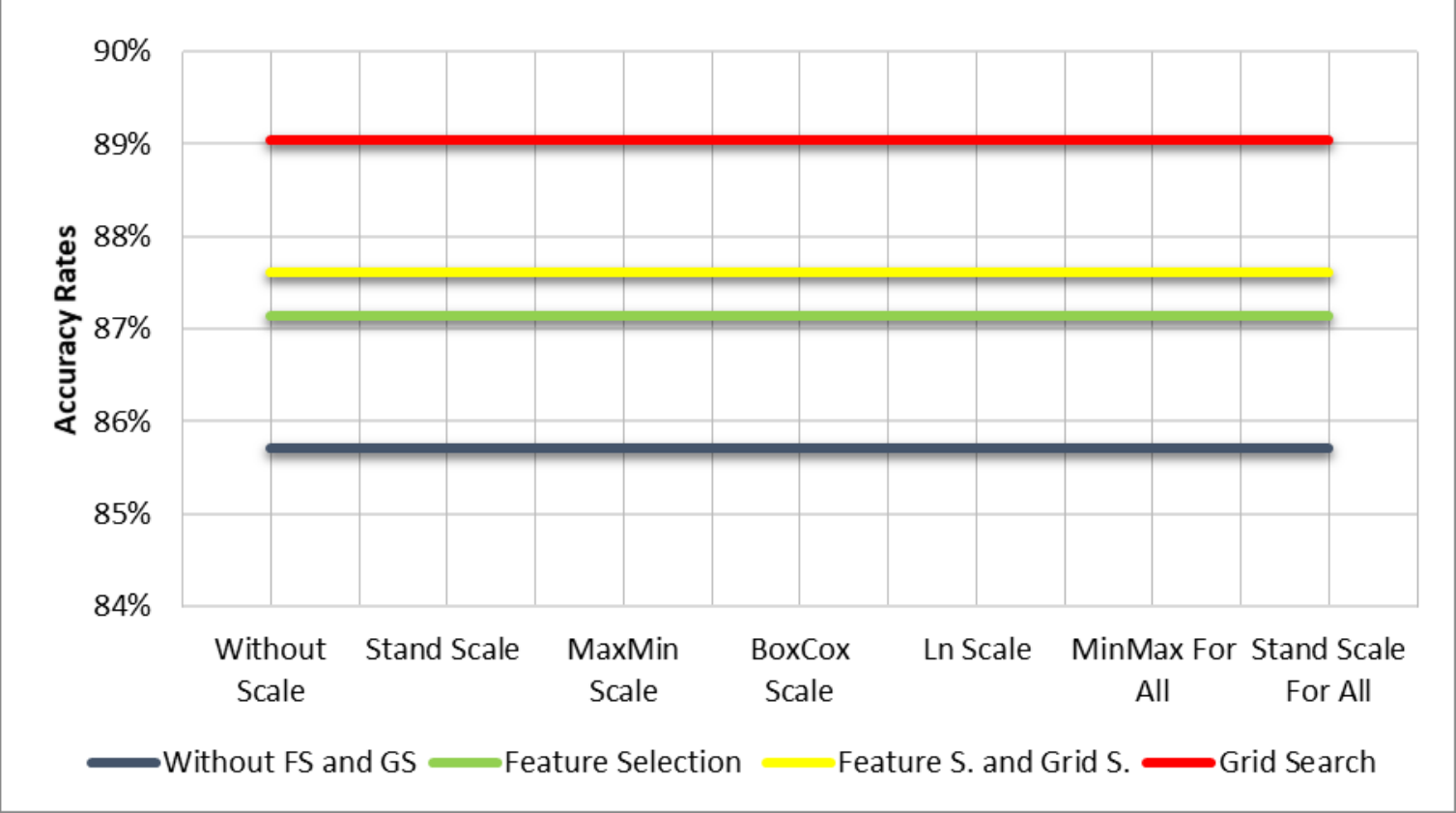}
				\label{fig:XGB}
			\end{subfigure} & \\
		\end{tabular}
		\label{fig:mytable}
\end{figure*}}

\end{document}